\documentstyle [12pt,eqsecnum,amsfonts,aps]{revtex}
\input epsf

\tighten
\draft
\widetext
\input epsf
\newcommand{\beq}{\begin{equation}}
\newcommand{\eeq}{\end{equation}}
\newcommand{\beqs}{\begin{eqnarray}}
\newcommand{\eeqs}{\end{eqnarray}}

\begin{document}
\draft
\baselineskip 6.0mm

\bigskip
\bigskip

\title{Transfer Matrices for the Partition Function of the Potts Model on
Cyclic and M\"obius Lattice Strips}

\bigskip

\author{
Shu-Chiuan Chang$^{a,b}$ \thanks{email: scchang@phys.ntu.edu.tw} \and
Robert Shrock$^{c}$ \thanks{email: robert.shrock@sunysb.edu}}

\bigskip

\address{(a) \ Department of Applied Physics, Faculty of Science \\
Tokyo University of Science \\
Tokyo 162-8601, Japan}

\address{(b) \ Physics Division \\
National Center for Theoretical Sciences at Taipei \\
National Taiwan University  \\
Taipei 10617, Taiwan }

\address{(c) \ C. N. Yang Institute for Theoretical Physics \\
State University of New York \\
Stony Brook, N. Y. 11794}

\bigskip

\maketitle

\bigskip

\begin{abstract}

We present a method for calculating transfer matrices for the $q$-state Potts
model partition functions $Z(G,q,v)$, for arbitrary $q$ and temperature
variable $v$, on cyclic and M\"obius strip graphs $G$ of the square (sq),
triangular (tri), and honeycomb (hc) lattices of width $L_y$ vertices and of
arbitrarily great length $L_x$ vertices.  For the cyclic case we express the
partition function as $Z(\Lambda,L_y \times L_x,q,v)=\sum_{d=0}^{L_y} c^{(d)}
Tr[(T_{Z,\Lambda,L_y,d})^m]$, where $\Lambda$ denotes lattice type, $c^{(d)}$
are specified polynomials of degree $d$ in $q$, $T_{Z,\Lambda,L_y,d}$ is the
transfer matrix in the degree-$d$ subspace, and $m=L_x$ ($L_x/2$) for
$\Lambda=sq, \ tri \ (hc)$, respectively. An analogous formula is given for
M\"obius strips.  We exhibit a method for calculating $T_{Z,\Lambda,L_y,d}$ for
arbitrary $L_y$.  Explicit results for arbitrary $L_y$ are given for
$T_{Z,\Lambda,L_y,d}$ with $d=L_y$ and $d=L_y-1$.  In particular, we find very
simple formulas the determinant $det(T_{Z,\Lambda,L_y,d})$, and trace
$Tr(T_{Z,\Lambda,L_y})$.  Corresponding results are given for the equivalent
Tutte polynomials for these lattice strips and illustrative examples are
included.  We also present formulas for self-dual cyclic strips of the square
lattice.

\end{abstract}

\bigskip
\bigskip

\newpage
\pagestyle{plain}
\pagenumbering{arabic}

\section{Introduction}

The $q$-state Potts model has served as a valuable model for the study of phase
transitions and critical phenomena \cite{potts,wurev}.  On a lattice, or, more
generally, on a (connected) graph $G$, at temperature $T$, this model is
defined by the partition function
\beq
Z(G,q,v) = \sum_{ \{ \sigma_n \} } e^{-\beta {\cal H}}
\label{zfun}
\eeq%
with the (zero-field) Hamiltonian
\beq
{\cal H} = -J \sum_{\langle i j \rangle} \delta_{\sigma_i \sigma_j}
\label{ham}
\eeq
where $\sigma_i=1,...,q$ are the spin variables on each vertex (site)
$i \in G$;
$\beta = (k_BT)^{-1}$; and $\langle i j \rangle$ denotes pairs of adjacent
vertices.  The graph $G=G(V,E)$ is defined by its vertex set $V$ and its edge
set $E$; we denote the number of vertices of $G$ as $n=n(G)=|V|$ and the
number of edges of $G$ as $e(G)=|E|$.  We use the notation
\beq
K = \beta J \ , \quad v = e^K - 1
\label{kdef}
\eeq
so that the physical ranges are $v \ge 0$ for the Potts ferromagnet, and $-1
\le v \le 0$ for Potts antiferromagnet, corresponding to $0 \le T \le \infty$.
One defines the (reduced) free energy per site $f=-\beta F$, where $F$ is the
actual free energy, via $f(\{G\},q,v) = \lim_{n \to \infty} \ln [
Z(G,q,v)^{1/n}]$, where we use the symbol $\{G\}$ to denote the formal limit
$\lim_{n \to \infty}G$ for a given family of graphs.  In the present context,
this $n \to \infty$ limit corresponds to the limit of infinite length for a
strip graph of the square lattice of fixed width and some prescribed boundary
conditions.

In this paper we shall present transfer matrices for the $q$-state Potts model
partition functions $Z(G,q,v)$, for arbitrary $q$ and temperature variable $v$,
on cyclic and M\"obius strip graphs $G$ of the square, triangular, and
honeycomb lattices of width $L_y$ vertices and of arbitrarily great length
$L_x$ vertices.  We label the lattice type as $\Lambda$ and abbreviate the
three respective types as $sq$, $tri$, and $hc$.  Each strip involves a
longitudinal repetition of $m$ copies of a particular subgraph.  For the
square-lattice strips, this is a column of squares.  It is convenient to
represent the strip of the triangular lattice as obtained from the
corresponding strip of the square lattice with additional diagonal edges
connecting, say, the upper-left to lower-right vertices in each square.  In
both these cases, the length is $L_x=m$ vertices.  We represent the strip of
the honeycomb lattice in the form of bricks oriented horizontally.  In this
case, since there are two vertices in 1-1 correspondence with each horizontal
side of a brick, $L_x=2m$ vertices.  Summarizing for all of three lattices, the
relation between the number of vertices and the number of repeated copies is
\beq
L_x=\cases{ m & if $\Lambda=sq$ \ or \ $tri$ \ or $G_D$ \cr
            2m & if $\Lambda=hc$ \cr }
\label{lxm}
\eeq
Here $G_D$ is the cyclic self-dual strip of the square lattice, to be discussed
further below.  For the cyclic case the partition function has the general form
\cite{a,cf}
\beq
Z(\Lambda, L_y \times L_x,cyc.,q,v) = \sum_{d=0}^{L_y} c^{(d)}
\sum_{j=1}^{n_Z(\Lambda,L_y,d)} (\lambda_{Z,\Lambda,L,d,j})^m
\label{zgsumcyc}
\eeq
In terms of transfer matrices, this can be written as 
\beq
Z(\Lambda, L_y \times L_x,cyc.,q,v) = 
\sum_{d=0}^{L_y} c^{(d)}Tr[(T_{Z,\Lambda,L_y,d})^m]
\label{zgsum_transfer}
\eeq
where $c^{(d)}$ are polynomials of degree $d$ in $q$ defined below, and the
transfer matrices in the degree-$d$ subspace $T_{Z,\Lambda,L_y,d}$ and their
eigenvalues $\lambda_{Z,\Lambda,L_y,d,j}$, $j=1,..., n_Z(\Lambda,L_y,d)$ are
independent of the length of the strip length $L_x$.  Here we shall construct
an analogous formula for M\"obius strips.  We exhibit our method for
calculating $T_{Z,\Lambda,L_y,d}$ for arbitrary $L_y$.  Explicit results for
arbitrary $L_y$ are given for (i) $T_{Z,\Lambda,L_y,d}$ with $d=L_y$ and
$d=L_y-1$, (ii) the determinant $det(T_{Z,\Lambda,L_y,d})$, and (iii) the trace
$Tr(T_{Z,\Lambda,L_y})$.  Corresponding results are given for the equivalent
Tutte polynomials for these lattice strips and illustrative examples are
included.  We have calculated the transfer matrices up to widths $L_y=5$ for
the square, triangular, and honeycomb lattices and $L_y=4$ for the cyclic
self-dual strip of the square lattice.  Since the dimensions of these matrices
increase rapidly with strip width (e.g., $dim(T_{Z,\Lambda,L_y,d})=14,28, 20,7$
for $L_y=4$ and $0 \le d \le 3$), it is not feasible to present many of the
explicit results here; instead, we concentrate on general methods and results
that hold for arbitrary $L_y$. 

Various special cases of the Potts model partition function $Z(G,q,v)$ are of
interest.  For example, if one considers the case of antiferromagnetic
spin-spin coupling, $J < 0$ and takes the temperature to zero, so that
$K=-\infty$ and $v=-1$, then
\beq
Z(G,q,-1)=P(G,q)
\label{zp}
\eeq
where $P(G,q)$ is the chromatic polynomial (in $q$) expressing the number of
ways of coloring the vertices of the graph $G$ with $q$ colors such that no two
adjacent vertices have the same color \cite{bbook,rtrev}.

We recall some previous related work.  The partition function $Z(G,q,v)$ for
the Potts model was calculated for arbitrary $q$ and $v$ on strips of the
lattice $\Lambda$ with cyclic or M\"obius longitudinal boundary conditions was
calculated for (i) $\Lambda=sq$, $L_y=2$ in \cite{a} and the nature of the
coefficients $c^{(d)}$ and subspace dimensions $n_Z(\Lambda,L_y,d)$ given in
\cite{cf}, (ii) $\Lambda=sq$, $L_y=3$ in \cite{s3a}, (iii) $\Lambda=tri$,
$L_y=3$ in \cite{ta}, (iv) for $\Lambda=hc$ in \cite{hca}, (v) for cyclic
self-dual strips of the square lattice with $L_y=1,2,3$ in \cite{jz,dg,sdg},
and (vi) for $L_y=2,3$ on the square-lattice with next-nearest-neighbor
spin-spin couplings in \cite{ka,ka3}.  Some general structural properties for
lattice strips with arbitrary width were given in \cite{cf,dg,s5}.  Matrix
methods for calculating chromatic polynomials were developed and used in
\cite{bds,bm}, \cite{baxter86,baxter87} and more recently in
\cite{matmeth}-\cite{matmeth3}.  Ref. \cite{sqtran} developed transfer matrix
methods for both $Z(G,q,v)$ and the special case $v=-1$ of chromatic
polynomials on strips of the square lattice with free longitudinal boundary
conditions and used them to calculate the latter polynomials for a large
variety of widths.  These have been termed transfer matrices in the
Fortuin-Kasteleyn representation (see eq. (\ref{cluster}) below).  These
methods were applied to calculate the full Potts model partition function for
strips of the square and triangular lattices with free boundary conditions and
a number of widths in Refs. \cite{ts,tt}.  A number of calculations have been
done for the special case of chromatic polynomials of lattice strip graphs; we
do not review these here but refer the reader to the references in \cite{pt}.
In a different direction, we mention calculations for arbitrary $q$ and $v$ on
rectangular lattice patches for the purpose of studying distributions of zeros
\cite{ks,ck}; we also do not review these studies here since our present focus
is calculations for strips of arbitrarily great length.

There are several motivations for presenting these transfer matrix methods for
calculating Potts model partition functions on cyclic and M\"obius lattice
strips and general results that we have obtained for lattice strips of
arbitrary width as well as length. Clearly, new exact results on the Potts
model are of value in their own right.  The transfer matrices are also a
convenient way to calculate partition functions.  In explaining this, one
should note that a very compact way of expressing the relevant information is
in the form of generating functions for $Z(G,q,v)$ or in the form where the
eigenvalues are given as roots of the characteristic polynomials.  For a
particular degree-$d$ subspace, it requires $n_Z(\Lambda,L_y,d)$ coefficients
to specify these characteristic polynomials.  In contrast, it requires
$n_Z(\Lambda,L_y,d)^2$ entries (some of which may be the same) to specify the
transfer matrix $T_{Z,\Lambda,L_y,d}$.  However, although the transfer matrix
is not the most compact way of presenting the necessary information to specify
$Z(G,q,v)$, it is, nevertheless, as convenient as the generating function
method for calculating the partition function since one only needs to compute
traces of powers of the matrices.  Moreover, except for the lowest few values
of the strip width $L_y$, the characteristic polynomials are often of fifth
order or higher, so that it is not possible to obtain explicit algebraic
solutions for the eigenvalues.  In our previous work we have made use of the
theorem on symmetric polynomial functions of roots of an algebraic equation
\cite{pm,uspensky}, which states that such functions can be expressed (via
Newton identities) in terms of the coefficients of the algebraic equation.  The
particular type of symmetric polynomial function of the roots that is relevant
for the Potts model partition functions on cyclic strips is the sum of $m$'th
powers of these roots.  Although the theorem on symmetric polynomial functions
of roots of algebraic equations guarantees that these sums of $m$'th powers are
expressible in terms of the coefficients of the equations, which are
polynomials in $q$ and $v$, it does not imply that they are particularly
simple, and, indeed, the Potts model partition functions of moderately long
lattice strips are quite lengthy expressions.  Since the determinant and trace
of a matrix are, respectively, the product and sum of the eigenvalues, they are
both symmetric polynomial function of these eigenvalues (roots of the
characteristic polynomial).  The well-known expressions for the determinant and
trace of a matrix in terms of coefficients of the characteristic polynomial are
examples of the above-mentioned theorem. The very simple formulas that we have
obtained for determinants and traces of transfer matrices provide further
motivation for the present exposition.

Let $G^\prime=(V,E^\prime)$ be a spanning subgraph of $G$, i.e. a subgraph
having the same vertex set $V$ and an edge set $E^\prime \subseteq E$. 
$Z(G,q,v)$ can be written as the sum \cite{kf,fk}
\beq
Z(G,q,v) = \sum_{G^\prime \subseteq G} q^{k(G^\prime)}v^{|E^\prime|}
\label{cluster}
\eeq
where $k(G^\prime)$ denotes the number of connected components of $G^\prime$.
Since we only consider connected graphs $G$, we have $k(G)=1$. The formula
(\ref{cluster}) enables one to generalize $q$ from ${\mathbb Z}_+$ to ${\mathbb
R}_+$ for physical ferromagnetic $v$.  More generally, eq. (\ref{cluster})
allows one to generalize both $q$ and $v$ to complex values, as is necessary
when studying zeros of the partition function in the complex $q$ and $v$
planes.  

The Potts model partition function $Z(G,q,v)$ is equivalent to an object of
considerable current interest in mathematical graph theory, the Tutte
polynomial, $T(G,x,y)$, given by \cite{tutte1}-\cite{tutte3}
\beq
T(G,x,y)=\sum_{G^\prime \subseteq G} (x-1)^{k(G^\prime)-k(G)}
(y-1)^{c(G^\prime)}
\label{tuttepol}
\eeq
where $c(G^\prime) = |E^\prime|+k(G^\prime)-|V|$ 
is the number of independent circuits in $G^\prime$.  Now let
\beq
x=1+\frac{q}{v}, \quad y=v+1
\label{xdef}
\eeq
so that
\beq
q=(x-1)(y-1) \ .  
\label{qxy}
\eeq
Then the equivalence between the Potts model partition function and the Tutte
polynomial for a graph $G$ is
\beq
Z(G,q,v)=(x-1)^{k(G)}(y-1)^{n(G)}T(G,x,y) \ .
\label{zt}
\eeq
Given this equivalence, we can express results either in Potts or Tutte form.
We will use both, since each has its own particular advantages.  The Potts
model form, involving the variables $q$ and $v$ is convenient for physical
applications, since $q$ specifies the number of states and determines the
universality class of the transition, and $v$ is the temperature variable.  The
Tutte form has the advantage that many expressions are simpler when written in
terms of the Tutte variables $x$ and $y$.  

We recall that for a planar graph $G=(V,E)$, one defines the (planar) dual
graph $G^*$ as the graph obtained by replacing each vertex (face) of $G$ by a
face (vertex) of $G^*$ and connecting the vertices of the resultant $G^*$ by
edges.  The graph $G$ is self-dual if and only if $G= G^*$.  For a planar graph
$G_{pl}$, it is evident from the definition (\ref{tuttepol}) that the Tutte
polynomial satisfies
\beq
T(G_{pl},x,y)=T(G_{pl}^*,y,x) \ . 
\label{tdual}
\eeq
Equivalently, 
\beq
Z(G_{pl},q,v) = v^{e(G_{pl})}q^{-c(G_{pl})} Z(G_{pl}^*,q,\frac{q}{v}) \ .
\label{zdual}
\eeq

The coefficients in eq. (\ref{zgsumcyc}) are 
\beq
c^{(d)} = U_{2d}(q^{1/2}/2) = \sum_{j=0}^d (-1)^j {2d-j \choose j}
q^{d-j}
\label{cd}
\eeq
with $U_n(x)$ being the Chebyshev polynomial of the second kind.  The first few
of these coefficients are $c^{(0)}=1$, $c^{(1)}=q-1$, $c^{(2)}=q^2-3q+1$, and
$c^{(3)}=q^3-5q^2+6q-1$.  The $c^{(d)}$'s play a role analogous to
multiplicities of eigenvalues $\lambda_{Z,\Lambda,L_y,d,j}$, although this
identification is formal, since $c^{(d)}$ may be zero or negative for the
physical values $q=1,2,3$ \cite{cf}.  For $q \ge 4$, $c^{(d)}$ is
positive-definite.  From (\ref{zt}), one can write the Tutte polynomial as 
\beq
T(\Lambda,L_y \times L_x,cyc.,x,y) = \frac{1}{x-1} \sum_{d=0}^{L_y} c^{(d)}
\sum_{j=1}^{n_Z(\Lambda,L_y,d)} (\lambda_{T,\Lambda,L_y,d,j})^m
\label{tgsumcyc}
\eeq
where $m$ is given in terms of $L_x$ by eq. (\ref{lxm}) and it is convenient to
factor out a factor of $1/(x-1)$. (This factor is always cancelled, since the
Tutte polynomial is a polynomial in $x$ as well as $y$.)  In terms of transfer
matrices,
\beq
T(\Lambda,L_y \times L_x,cyc.,x,y) = 
\frac{1}{x-1}\sum_{d=0}^{L_y} c^{(d)} Tr[(T_{T,\Lambda,L_y,d})^m] \ . 
\label{tgsum_transfer}
\eeq
The equivalence of eq. (\ref{tgsum_transfer}) and (\ref{zgsum_transfer}) to
(\ref{tgsumcyc}) and (\ref{zgsumcyc}), respectively, relies upon the theorem
that an arbitrary square matrix $T$ can be put into (upper or lower) triangular
form (tf) by a (unitary) similarity transformation \cite{triform}
\beq
U T U^{-1} = T_{tf} \ . 
\label{ata}
\eeq
For definiteness, we consider the upper triangular form, for which
$(T_{tf})_{ij}=0$ if $i > j$. A matrix in triangular form has its eigenvalues
$\lambda_j$ as its diagonal elements and has the properties that (i) If $S$ and
$T$ are upper triangular matrices with eigenvalues $\lambda_{S,j}$ and
$\lambda_{T,j}$, then $ST$ is also a triangular matrix, with diagonal
elements $(ST)_{jj} = \lambda_{S,j}\lambda_{T,j}$.  A corollary is that for an
arbitrary $N \times N$ matrix $T$
\beq
Tr(T^m) = Tr[(T_{tf})^m] = \sum_{j=1}^N \lambda_j^m \ . 
\label{traceT}
\eeq

The dimension of $T_{Z,\Lambda,L_y,d}$, or equivalently, 
$T_{T,\Lambda,L_y,d}$, is \cite{cf}
\beq
n_Z(\Lambda,L_y,d)=n_T(\Lambda,L_y,d)=
\frac{(2d+1)}{(L_y+d+1)}{2L_y \choose L_y-d} \quad {\rm for} \ \
\Lambda=sq,tri,hc \ . 
\label{nzlyd}
\eeq
for $0 \le d \le L_y$ and zero otherwise.  
The property that the numbers $n_Z(\Lambda,L_y,d)$ are the same
for all three lattices $\Lambda=sq,tri,hc$, as was shown for $\Lambda=sq,tri$
in Ref. \cite{cf} and for $\Lambda=hc$ in \cite{hca}.  For this reason we shall
henceforth usually revert to our previous notation in Ref. \cite{cf}, setting
\beq
n_Z(\Lambda,L_y,d) \equiv n_Z(L_y,d) \ , \quad \Lambda=sq,tri,hc
\label{nznotation}
\eeq
The formal quantity $n_Z(0,d)$ will appear in some determinant formulas below
and is given by eq. (\ref{nzlyd}) as the Kronecker delta,
$n_Z(0,d)=\delta_{d,0}$.  (Below we shall consider self-dual strips $G_D$ of
the square lattice, which have different dimensions $n_Z(G_D,L_y,d)$; for this
case we shall include the $G_D$ dependence in the notation.) Special cases of
$n_Z(L_y,d)$ that are of interest here include
\beq
n_Z(L_y,L_y)=1
\label{nzlyly}
\eeq
\beq
n_Z(L_y,L_y-1)=2L_y-1
\label{nzlylyminus1}
\eeq
\beq
n_Z(L_y,0)=C_{L_y}
\label{nzly0}
\eeq
where here $C_n$ is the Catalan number which occurs in combinatorics and is 
defined by
\beq
C_n = \frac{1}{(n+1)}{2n \choose n} \ . 
\label{catalan}
\eeq
(No confusion should result from our use of the same symbol $C_n$ to denote 
the circuit graph with $n$ vertices since the meaning will be clear
from context).  The first few Catalan numbers are $C_1=1$, $C_2=2$, $C_3=5$, 
$C_4=14$, and $C_5=42$.

The full transfer matrix $T_{X,\Lambda,L_y}$, $X=Z,T$, has a block structure
formally specified by
\beq
T_{X,\Lambda,L_y} = \bigoplus_{d=0}^{L_y} \prod
T_{X,\Lambda,L_y,d} \quad X=Z,T
\label{Tdirectsum}
\eeq
where the product $\prod T_{X,\Lambda,L_y,d}$ means a set of square blocks,
each of dimension, $c^{(d)}$, of the form $\lambda_{X,\Lambda,L_y,d,j}$ times
the identity matrix. The dimension of the total transfer matrix, i.e., the
total number of eigenvalues $\lambda_{X,\Lambda,L_y,d,j}$, 
counting multiplicities, is thus
\beq
dim(T_{X,\Lambda,L_y}) = \sum_{d=0}^{L_y}
dim(T_{X,\Lambda,L_y,d}) = \sum_{d=0}^{L_y} c^{(d)}
n_Z(\Lambda,L_y,d) \quad X=Z,T\ .
\label{dimTL}
\eeq
As in our earlier work, we define $N_{Z,\Lambda,L_y,\lambda}$ as the total
number of distinct eigenvalues of $T_{X,\Lambda,L_y}$, $X=Z,T$, i.e.  the sum
of the dimensions of the submatrices $T_{X,\Lambda,L_y,d}$, modulo the
multiplicity $c^{(d)}$.  This is \cite{cf}
\beq
N_{Z,\Lambda,L_y}=N_{T,\Lambda,L_y}=\sum_{d=0}^{L_y} n_Z(L_y,d) = 
{2L_y \choose L} \quad {\rm for} \ \ \Lambda=sq,tri,hc \ . 
\label{nztot}
\eeq

\section{Transfer Matrix Method}

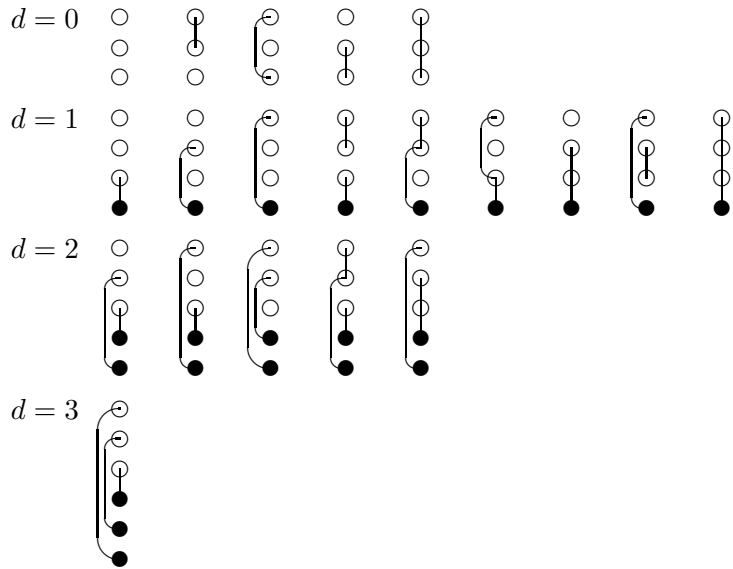
\begin{figure}
\unitlength 1mm \hspace*{5mm}
\begin{picture}(60,8)
\put(10,8){\makebox(0,0){{\small $d=0$}}}
\multiput(20,0)(10,0){5}{\circle{2}}
\multiput(20,4)(10,0){5}{\circle{2}}
\multiput(20,8)(10,0){5}{\circle{2}} \put(30,4){\line(0,1){4}}
\put(40,4){\oval(4,8)[l]} \put(50,0){\line(0,1){4}}
\put(60,0){\line(0,1){8}}
\end{picture}

\vspace*{5mm} \hspace*{5mm}
\begin{picture}(100,12)
\put(10,12){\makebox(0,0){{\small $d=1$}}}
\multiput(20,0)(10,0){9}{\circle*{2}}
\multiput(20,4)(10,0){9}{\circle{2}}
\multiput(20,8)(10,0){9}{\circle{2}}
\multiput(20,12)(10,0){9}{\circle{2}} \put(20,0){\line(0,1){4}}
\put(30,4){\oval(4,8)[l]} \put(40,6){\oval(4,12)[l]}
\put(50,0){\line(0,1){4}} \put(50,8){\line(0,1){4}}
\put(60,4){\oval(4,8)[l]} \put(60,8){\line(0,1){4}}
\put(70,0){\line(0,1){4}} \put(70,8){\oval(4,8)[l]}
\put(80,0){\line(0,1){8}} \put(90,6){\oval(4,12)[l]}
\put(90,4){\line(0,1){4}} \put(100,0){\line(0,1){12}}
\end{picture}

\vspace*{5mm} \hspace*{5mm}
\begin{picture}(60,16)
\put(10,16){\makebox(0,0){{\small $d=2$}}}
\multiput(20,0)(10,0){5}{\circle*{2}}
\multiput(20,4)(10,0){5}{\circle*{2}}
\multiput(20,8)(10,0){5}{\circle{2}}
\multiput(20,12)(10,0){5}{\circle{2}}
\multiput(20,16)(10,0){5}{\circle{2}} \put(20,4){\line(0,1){4}}
\put(20,6){\oval(4,12)[l]} \put(30,4){\line(0,1){4}}
\put(30,8){\oval(4,16)[l]} \put(40,8){\oval(4,8)[l]}
\put(40,8){\oval(6,16)[l]} \put(50,4){\line(0,1){4}}
\put(50,6){\oval(4,12)[l]} \put(50,12){\line(0,1){4}}
\put(60,4){\line(0,1){8}} \put(60,8){\oval(4,16)[l]}
\end{picture}

\vspace*{5mm} \hspace*{5mm}
\begin{picture}(20,20)
\put(10,20){\makebox(0,0){{\small $d=3$}}}
\multiput(20,0)(0,4){3}{\circle*{2}}
\multiput(20,12)(0,4){3}{\circle{2}} \put(20,8){\line(0,1){4}}
\put(20,10){\oval(4,12)[l]} \put(20,10){\oval(6,20)[l]}
\end{picture}

\caption{\footnotesize{Partitions for the $L_y=3$ strip.}}
\label{L3partitions}
\end{figure}

There are several equivalent ways to calculate $Z(G,q,v)$ or $T(G,x,y)$ for
these lattice strip graphs.  One is to make iterative use of the
deletion-contraction relations obeyed by $T(G,x,y)$.  This yields a generating
function whose denominator directly determines the
$\lambda_{X,\Lambda,L_y,d,j}$'s where $X=Z,T$.  This iterative method works
also for strip graphs with free longitudinal boundary conditions.  The transfer
matrix method for lattice strips of the square and triangular lattices with
free longitudinal boundary conditions (and free or cylindrical transverse
boundary conditions) was explained in detail in Ref. \cite{sqtran}.  For the
free strip with width $L_y$, the bases of the transfer matrix are all of the
possible non-crossing partitions of $L_y$ vertices.  (For the zero-temperature
antiferromagnetic Potts model, the non-nearest-neighbor requirement is
imposed.)  The eigenvalues of the transfer matrix for a free strip are the same
as the eigenvalues of the transfer matrix $T_{Z,\Lambda,L_y,d=0}$ in the degree
$d=0$ subspace (called ``level 0'' subspace in \cite{matmeth3}) for the
corresponding cyclic strip, and the dimension of this matrix is
$n_Z(L_y,0)=C_{L_y}$, the Catalan number, as in eq. (\ref{nzly0}). This was
shown for the free strips of the square and triangular lattices in
\cite{sqtran} and for the cyclic/M\"obius strips in \cite{cf} and extended to
the honeycomb lattice in \cite{hca}.  As discussed in
Refs. \cite{matmeth}-\cite{matmeth3} for chromatic polynomials, which we
generalize here to the full Potts model partition function, the degree $d=1$
subspace is given by all of the possible non-crossing partitions with a color
assignment to one vertex (with possible connections with other vertices), and
the multiplicity is $q-1=c^{(1)}$. This follows because there are $q$ possible
ways of making this color assignment, but one of these has to be subtracted,
since the effect of all the possible color assignments is equivalent to the
choice of no specific color assignment, which has been taken into account in
the level 0 subspace.  In this derivation and subsequent ones we assume that
$q$ is an integer $\ge 4$ to begin with, so that the multiplicities are
positive-definite; we then analytically continue them downward to apply in the
region $0 \le q < 4$ where $c^{(d)}$ can be zero or negative.  For the next
subspace we consider all of the non-crossing partitions with two-color
assignments to two separated vertices (with possible connections with other
vertices). Now the multiplicity can be understood by the sieve formula of
\cite{matmeth}-\cite{matmeth3}. Since the two assigned colors should be
different, there are $q(q-1)$ ways of making these assignments. This includes
the $q$ possible color assignments for each of the two vertices that have been
considered in level 1 and hence these must be subtracted. In doing this, the
no-color assignment was subtracted twice, and one of these has to be added
back. Therefore, the multiplicity is
\beq
q(q-1)-2q+1 = q^2-3q+1 = c^{(2)} \ .
\label{cd2derive}
\eeq
By the same method, for the non-crossing partitions with the three-color
assignment, the multiplicity is
\beq
q(q-1)^2-2q(q-1)-q^2+3q-1 = q^3-5q^2+6q-1 = c^{(3)} \ . 
\label{cd3derive}
\eeq
The multiplicity for the four-color assignment is
\beqs
& & q(q-1)^3-2q(q-1)^2-2q^2(q-1)+3q(q-1)+3q^2-4q+1 \cr\cr
& = & (q-1)(q^3-6q^2+9q-1) = c^{(4)} \ . 
\label{cd4derive}
\eeqs
and so forth for higher values of $d$.  We show the calculation for these
multiplicities pictorially for $2 \le d \le 4$ in Fig. \ref{cdfigure}. In
general, the multiplicity of the $d$-color assignment can be computed to be
$c^{(d)}$ given in eq. (\ref{cd}). We list graphically all the possible
partitions for $L_y=2$ and $L_y=3$ strips in Figs. \ref{L2partitions} and
\ref{L3partitions}, respectively, where white circles are the original $L_y$
vertices and each black circle corresponds to a specific color assignment. We
remark that the connections between the black circles and white circles also
obey the non-crossing restriction, so it is possible that two non-adjacent
black circles represent the same color. In the following discussion, we will
simply use the names white and black circles with the meaning understood.  We
denote the partitions ${\cal P}_{L_y,d}$ for $2 \le L_y \le 4$ as follows:
\beq 
{\cal P}_{2,0} = \{ I; 12 \} \ , \qquad {\cal P}_{2,1} = \{ \bar
2; \bar 1; \overline{12} \} \ , \qquad {\cal P}_{2,2} = \{ \bar 1, \bar 2 \}
\label{L2partitionlist} 
\eeq
\beqs 
{\cal P}_{3,0} & = & \{ I; 12; 13; 23; 123 \} \ , \qquad {\cal P}_{3,1} =
\{ \bar 3; \bar 2; \bar 1; 12, \bar 3; \overline{12}; \overline{13};
\overline{23}; 23, \bar 1; \overline{123} \} \ , \cr\cr {\cal P}_{3,2} & = & \{
\bar 2, \bar 3; \bar 1, \bar 3; \bar 1, \bar 2; \overline{12}, \bar 3; \bar 1,
\overline{23} \} \ , \qquad {\cal P}_{3,3} = \{ \bar 1, \bar 2, \bar 3 \}
\label{L3partitionlist} 
\eeqs
\beqs 
{\cal P}_{4,0} & = & \{ I; 12; 13; 14; 23; 24; 34; 12, 34; 14, 23; 123;
124; 134; 234; 1234 \} \ , \cr\cr {\cal P}_{4,1} & = & \{ \bar 4; \bar 3; \bar
2; \bar 1; 12, \bar 4; 12, \bar 3; \overline{12}; 13, \bar 4; \overline{13};
\overline{14}; 23, \bar 4; \overline{23}; 23, \bar 1; \overline{24}; 24, \bar
1; \overline{34}; 34, \bar 2; 34, \bar 1; 12, \overline{34}; \cr\cr & & 34,
\overline{12}; 23, \overline{14}; 123, \bar 4; \overline{123}; \overline{124};
\overline{134}; \overline{234}; 234, \bar 1; \overline{1234} \} \ , \cr\cr
{\cal P}_{4,2} & = & \{ \bar 3, \bar 4; \bar 2, \bar 4; \bar 1, \bar 4; \bar 2,
\bar 3; \bar 1, \bar 3; \bar 1, \bar 2; 12, \bar 3, \bar 4; \overline{12}, \bar
4; \overline{12}, \bar 3; \overline{13}, \bar 4; \overline{23}, \bar 4; 23,
\bar 1, \bar 4; \bar 1, \overline{23}; \bar 1, \overline{24}; \bar 2,
\overline{34}; \cr\cr & & \bar 1, \overline{34}; 34, \bar 1, \bar 2;
\overline{12}, \overline{34}; \overline{123}, \bar 4; \bar 1, \overline{234} \}
\ , \cr\cr {\cal P}_{4,3} & = & \{ \bar 2, \bar 3, \bar 4; \bar 1, \bar 3, \bar
4; \bar 1, \bar 2, \bar 4; \bar 1, \bar 2, \bar 3; \overline{12}, \bar 3, \bar
4; \bar 1, \overline{23}, \bar 4; \bar 1, \bar 2, \overline{34} \} \ , \cr\cr
{\cal P}_{4,4} & = & \{ \bar 1, \bar 2, \bar 3, \bar 4 \}
\label{L4partitionlist} 
\eeqs
\begin{figure}
\unitlength 1mm \hspace*{5mm}
\begin{picture}(135,8)
\put(10,8){\makebox(0,0){{\small $d=2$}}}
\multiput(20,4)(0,4){2}{\circle{2}} \put(20,4){\line(0,1){4}}
\put(25,6){\makebox(0,0){{\small $-$}}}
\put(30,6){\makebox(0,0){{\small $2$}}} \put(35,6){\circle{2}}
\put(40,6){\makebox(0,0){{\small $+$}}}
\put(45,6){\makebox(0,0){{\small $1$}}}
\put(65,8){\makebox(0,0){{\small $d=3$}}}
\multiput(75,0)(0,4){3}{\circle{2}} \put(75,0){\line(0,1){8}}
\put(80,4){\makebox(0,0){{\small $-$}}}
\put(85,4){\makebox(0,0){{\small $($}}}
\put(90,4){\makebox(0,0){{\small $2$}}}
\multiput(95,2)(0,4){2}{\circle{2}} \put(95,2){\line(0,1){4}}
\put(100,4){\makebox(0,0){{\small $+$}}}
\multiput(105,2)(0,4){2}{\circle{2}}
\put(110,4){\makebox(0,0){{\small $)$}}}
\put(115,4){\makebox(0,0){{\small $+$}}}
\put(120,4){\makebox(0,0){{\small $3$}}} \put(125,4){\circle{2}}
\put(130,4){\makebox(0,0){{\small $-$}}}
\put(135,4){\makebox(0,0){{\small $1$}}}
\end{picture}

\vspace*{5mm} \hspace*{5mm}
\begin{picture}(125,12)
\put(10,12){\makebox(0,0){{\small $d=4$}}}
\multiput(20,0)(0,4){4}{\circle{2}} \put(20,0){\line(0,1){12}}
\put(25,6){\makebox(0,0){{\small $-$}}}
\put(30,6){\makebox(0,0){{\small $($}}}
\put(35,6){\makebox(0,0){{\small $2$}}}
\multiput(40,2)(0,4){3}{\circle{2}} \put(40,2){\line(0,1){8}}
\put(45,6){\makebox(0,0){{\small $+$}}}
\put(50,6){\makebox(0,0){{\small $2$}}}
\multiput(55,2)(0,4){3}{\circle{2}} \put(55,6){\line(0,1){4}}
\put(60,6){\makebox(0,0){{\small $)$}}}
\put(65,6){\makebox(0,0){{\small $+$}}}
\put(70,6){\makebox(0,0){{\small $($}}}
\put(75,6){\makebox(0,0){{\small $3$}}}
\multiput(80,4)(0,4){2}{\circle{2}} \put(80,4){\line(0,1){4}}
\put(85,6){\makebox(0,0){{\small $+$}}}
\put(90,6){\makebox(0,0){{\small $3$}}}
\multiput(95,4)(0,4){2}{\circle{2}}
\put(100,6){\makebox(0,0){{\small $)$}}}
\put(105,6){\makebox(0,0){{\small $-$}}}
\put(110,6){\makebox(0,0){{\small $4$}}} \put(115,6){\circle{2}}
\put(120,6){\makebox(0,0){{\small $+$}}}
\put(125,6){\makebox(0,0){{\small $1$}}}
\end{picture}

\caption{\footnotesize{Multiplicities of the transfer matrices for
$2 \le d \le 4$.}} \label{cdfigure}
\end{figure}

\begin{figure}
\unitlength 1mm \hspace*{5mm}
\begin{picture}(110,12)
\put(10,12){\makebox(0,0){{\small $d=0$}}}
\multiput(20,8)(10,0){2}{\circle{2}}
\multiput(20,12)(10,0){2}{\circle{2}} \put(30,8){\line(0,1){4}}
\put(50,12){\makebox(0,0){{\small $d=1$}}}
\multiput(60,4)(10,0){3}{\circle*{2}}
\multiput(60,8)(10,0){3}{\circle{2}}
\multiput(60,12)(10,0){3}{\circle{2}} \put(60,4){\line(0,1){4}}
\put(70,8){\oval(4,8)[l]} \put(80,4){\line(0,1){8}}
\put(100,12){\makebox(0,0){{\small $d=2$}}}
\multiput(110,0)(0,4){2}{\circle*{2}}
\multiput(110,8)(0,4){2}{\circle{2}} \put(110,4){\line(0,1){4}}
\put(110,6){\oval(4,12)[l]}
\end{picture}

\caption{\footnotesize{Partitions for the $L_y=2$ strip.}}
\label{L2partitions}
\end{figure}

Using these methods, we can obtain $n_Z(L_y,d)$ given in Theorem 4 of \cite{cf}
as follows. Since the maximum number of colors to assign is $L_y$ for a strip
with width $L_y$, it follows that $n_Z(L_y,d)=0$ for $d>L_y$ and
$n_Z(L_y,L_y)=1$. It is elementary that $n_Z(1,0)=1$. The $d=0$ partitions of a
width-$(L_y+1)$ strip can be obtained by either adding a unconnected white circle
to the bottom of the $d=0$ partitions of a width-$L_y$ strip or converting
the black circle of the $d=1$ partitions of a width-$L_y$ strip into a white
circle.  That is, $n_Z(L_y+1,0) = n_Z(L_y,0) + n_Z(L_y,1)$. Finally, the
partitions of a width-$(L_y+1)$ strip for $1 \le d \le L_y+1$ can be obtained
in one of the following four ways: (a) adding a pair of connected circles, one
black and one white, (but not connected to any other vertex) above the highest
black circle of the $d-1$ partitions of a width-$L_y$ strip; (b) adding a
unconnected white circle above the highest black circle of the $d$ partitions
of a width-$L_y$ strip; (c) adding a white circle above the highest black
circle of the $d$ partitions of a width-$L_y$ strip and connecting these two
circles; (d) converting the highest black circle of the $d+1$ partitions of a
width-$L_y$ strip into a white circle. Now the lowest white circle of the $d$
partitions of a width-$(L_y+1)$ strip can either connect to a black circle
with or without ((c) or (a)) other connections to other white circles, or it
does not connect to a black circle with or without ((d) or (b)) other
connections to other white circles. Therefore, (a) to (d) exhaust all the
possibilities, and we have $n_Z(L_y+1,d) = n_Z(L_y,d-1) + 2n_Z(L_y,d) +
n_Z(L_y,d+1)$ for $1 \le d \le L_y+1$. This completes the proof.  Indeed, these
are precisely the equations that we obtained in a different manner in
Ref. \cite{cf}.  Solving these equations as we did in \cite{cf}, it follows
that the dimension of the transfer matrix $T_{Z,\Lambda,L_y,d}$ is the
expression for $n_Z(L_y,d)$ given in eq. (\ref{nzlyd}).

Similarly, for the zero-temperature antiferromagnetic Potts model, $n_P(\Lambda,L_y,d)$ for $\Lambda = sq,tri$
given in Theorem 2 of \cite{cf} can be shown as follows. Now the adjacent white
circles cannot connect to each other. Again, it is obvious that $n_P(L_y,d)=0$
for $d>L_y$, $n_P(L_y,L_y)=1$, and $n_P(1,0)=1$. The $d=0$ partitions of a
width-$(L_y+1)$ strip can all be obtained by converting the black circle of
the $d=1$ partitions of a width-$L_y$ strip into a white circle with the
condition that the connection between the black circle and the lowest white
circle, if there is one, must be deleted.  That is, $n_P(L_y+1,0) =
n_P(L_y,1)$. Finally, the partitions of a width-$(L_y+1)$ strip for $1 \le d
\le L+1$ can be obtained in one of the following four ways: (a) adding a pair
of connected circles, one black and one white, (but not connected to any other
vertex) above the highest black circle of the $d-1$ partitions of a width-$L_y$ strip; (b) adding a unconnected white circle above the highest black
circle of the $d$ partitions of a width-$L_y$ strip; (c) if the highest
black circle of the $d+1$ partitions of a width-$L_y$ strip connects to the
lowest white circle, deleting this connection, converting this black circle
into a white circle, and connecting it to the black circle below it; (d) if the highest black
circle of the $d+1$ partitions of a width-$L_y$ strip does not connect to the
lowest white circle, converting it into a white circle. Now the
lowest white circle of the $d$ partitions of a width-$(L_y+1)$ strip can either
connect to a black circle with or without ((c) or (a)) other non-adjacent connections to
other white circles, or it does not connect to a black circle with or without
((d) and (b)) other non-adjacent connections to other white circles. Therefore, (a) to (d)
exhaust all the possibilities, and we have $n_P(L_y+1,d) = n_P(L_y,d-1) +
n_P(L_y,d) + n_P(L_y,d+1)$ for $1 \le d \le L+1$. This completes the proof.

The construction of the transfer matrix for each level (= degree) $d$ can be
carried out by the methods similar to those for $d=0$. For cyclic strips, the
transfer matrix $T_{Z,\Lambda,L_y,d}$ is the product of the transverse and
longitudinal parts, $H_{Z,\Lambda,L_y,d}$ and $V_{Z,\Lambda,L_y,d}$, which can
be expressed as
\beqs 
H_{Z,sq,L_y,d} & = & H_{Z,tri,L_y,d} = \prod_{i=1}^{L_y-1}
(I+vJ_{L_y,d,i,i+1}) \cr\cr H_{Z,hc,L_y,d,1} & = & \prod
_{i=1}^{[L_y/2]} (I+vJ_{L_y,d,2i-1,2i}) \ , \qquad H_{Z,hc,L_y,d,2} =
\prod _{i=1}^{[(L_y-1)/2]} (I+vJ_{L_y,d,2i,2i+1}) \cr\cr V_{Z,sq,L_y,d}
& = & V_{Z,hc,L_y,d} = \prod_{i=1}^{L_y} (vI+D_{L_y,d,i}) \cr\cr
V_{Z,tri,L_y,d} & = & \prod_{i=1}^{L_y-1} [(vI+D_{L_y,d,i})
(I+vJ_{L_y,d,i,i+1})] (vI+D_{L_y,d,L_y}) \ , 
\label{HVmatrix} 
\eeqs
where $[\nu]$ denotes the integral part of $\nu$, and $I$ is the identity
matrix of dimension $n_Z(L_y,d)$. $J_{L_y,d,i,i+1}$ is the join operator
between vertices $i$ and $i+1$, and $D_{L_y,d,i}$ is the detach
operator on vertex $i$ \cite{sqtran}. We have
\beqs 
T_{Z,sq,L_y,d} & = & V_{Z,sq,L_y,d} H_{Z,sq,L_y,d} \ , \qquad
T_{Z,tri,L_y,d} = V_{Z,tri,L_y,d} H_{Z,tri,L_y,d} \cr\cr T_{Z,hc,L_y,d} &
= & (V_{Z,hc,L_y,d} H_{Z,hc,L_y,d,2}) (V_{Z,hc,L_y,d} H_{Z,hc,L_y,d,1})
\equiv T_{Z,hc,L_y,d,2} T_{Z,hc,L_y,d,1} \ . 
\label{transfermatrix}
\eeqs
Here we shall give results for this method for both cyclic and M\"obius strips.

Consider two sets of $L_y$ vertices and denote them as $i_1, i_2, ... i_{L_y}$
and $j_1, j_2, ... j_{L_y}$. For cyclic strips of the square lattice, the
horizontal edges connecting these vertices are $(i_1,j_1)$, $(i_2,j_2)$, ...,
$(i_{L_y},j_{L_y})$. For M\"obius strips, one set of horizontal edges becomes
$(i_1,j_{L_y})$, $(i_2,j_{L_y-1})$, ..., $(i_{L_y},j_1)$. This corresponds to
exchanging the pair of bases that switch to each other when the vertices
reverse in order, i.e., the set of bases that do not have self-reflection
symmetry with respect to the center of the $L_y$ vertices. For example, among
the partitions for the $L_y=2$ strip in Fig. \ref{L2partitions}, the first
partition $\bar 2$ and the second partition $\bar 1$ in ${\cal P}_{2,1}$ must
be exchanged under this reflection. The pairs of partitions for the $L_y=3$
strip in Fig. \ref{L3partitions} are (1) the second partition $12$ and the
fourth partition $23$ in ${\cal P}_{3,0}$, (2) the first partition $\bar 3$ and
the third partition $\bar 1$ in ${\cal P}_{3,1}$, (3) the fifth partition
$\overline{12}$ and the seventh partition $\overline{23}$ in ${\cal P}_{3,1}$,
(4) the fourth partition $12, \bar 3$ and the eighth partition $23, \bar 1$ in
${\cal P}_{3,1}$, (5) the first partition $\bar 2, \bar 3$ and the third
partition $\bar 1, \bar 2$ in ${\cal P}_{3,2}$, and (6) the fourth partition
$\overline{12}, \bar 3$ and the fifth partition $\bar 1, \overline{23}$ in
${\cal P}_{3,2}$. For this specific set of edges of M\"obius strips, the pairs
of columns of $V_{Z,\Lambda,L_y,d}$ that correspond to these pairs of
partitions should be exchanged, and these matrices will be denoted as $\tilde
V_{Z,\Lambda,L_y,d}$. Equivalently, the same pairs of columns of
$T_{Z,\Lambda,L_y,d}$ should be exchanged, and these matrices will be denoted
as $\tilde T_{Z,\Lambda,L_y,d}=\tilde V_{Z,\Lambda,L_y,d} H_{Z,\Lambda,L_y,d}$
for $\Lambda = sq, tri$.  There are two kinds of M\"obius strips for the
honeycomb lattice.  When $L_y$ is even, the number of vertices in the
horizontal direction is even as for the cyclic strips, i.e., $L_x = 2m$. When
$L_y$ is odd, the number of vertices in the horizontal direction is odd, $L_x =
2m-1$. Therefore, for the honeycomb lattice, we use the definition
\beqs 
\tilde T_{Z,hc,L_y,d} & = & \tilde V_{Z,hc,L_y,d} H_{Z,hc,L_y,d,1}
\qquad \mbox{for odd} \ L_y \cr\cr \tilde T_{Z,hc,L_y,d} & = & \tilde
V_{Z,hc,L_y,d} H_{Z,hc,L_y,d,2} V_{Z,hc,L_y,d} H_{Z,hc,L_y,d,1} \qquad
\mbox{for even} \ L_y \ . 
\eeqs
As was discussed in \cite{hca} for the crossing-subgraph strips, the square of
each eigenvalue of $\tilde T_{Z,hc,L_y,d}$ for odd $L_y$ is an eigenvalue of
$T_{Z,hc,L_y,d}$.

We have found in \cite{cf} the changes of coefficients for the square lattice
when the longitudinal boundary condition is changed from cyclic to M\"obius.
In the context of our present transfer matrix formalism, we can express these
changes of coefficients for the square, triangular and honeycomb lattices as
follows: 
\beq 
c^{(0)} \to c^{(0)} 
\label{cd0tran} 
\eeq
\beq 
c^{(2k)} \to -c^{(k-1)} \ , \qquad 1 \le k \le \Bigl [
\frac{L_y}{2} \Bigr ] 
\label{cdeventran} 
\eeq
\beq 
c^{(2k+1)} \to c^{(k+1)} \ ,  \qquad 0 \le k \le \Bigl [
\frac{L_y-1}{2} \Bigr ] \ . 
\label{cdoddtran} 
\eeq
The Potts model partition function for M\"obius strips is given by
\beqs 
Z(\Lambda,L \times L_x, Mb,q,v) & = &
c^{(0)}Tr[(T_{Z,\Lambda,L_y,0})^{m-1} \tilde T_{Z,\Lambda,L_y,0}] \cr\cr & & +
\sum_{d=0}^{[(L_y-1)/2]} c^{(d+1)} Tr[(T_{Z,\Lambda,L_y,2d+1})^{m-1}
\tilde T_{Z,\Lambda,L_y,2d+1}] \cr\cr & & - \sum_{d=1}^{[L_y/2]}
c^{(d-1)} Tr[(T_{Z,\Lambda,L_y,2d})^{m-1} \tilde T_{Z,\Lambda,L_y,2d}]
\label{zgsum_transfermb} 
\eeqs
The corresponding formula for the Tutte polynomial is
\beqs T(\Lambda,L \times L_x, Mb,x,y) & = & \frac{1}{x-1} \Bigl
( c^{(0} Tr[(T_{T,\Lambda,L_y,0})^{m-1} \tilde T_{T,\Lambda,L_y,0}] \cr\cr & &
+ \sum_{d=0}^{[(L_y-1)/2]} c^{(d+1)} Tr[(T_{T,\Lambda,L_y,2d+1})^{m-1}
\tilde T_{T,\Lambda,L_y,2d+1}] \cr\cr & & - \sum_{d=1}^{[L_y/2]}
c^{(d-1)} Tr[(T_{T,\Lambda,L_y,2d})^{m-1} \tilde T_{T,\Lambda,L_y,2d}]\Bigr ) 
\ . 
\label{tgsum_transfermb} 
\eeqs

For the square lattice or the honeycomb lattice with $L_y$ even, the
eigenvalues of $\tilde T_{Z,\Lambda,L_y,d}$ are the same as those of
$T_{Z,\Lambda,L_y,d}$ except for possible changes of signs. The number of
eigenvalues with sign changes is equal to the number of column-exchanges from
$T_{Z,\Lambda,L_y,d}$ to $\tilde T_{Z,\Lambda,L_y,d}$. Denote the number of
eigenvalues that are the same for $T_{Z,sq,L_y,d}$ and $\tilde 
T_{Z,sq,,L_y,d}$ as $n_Z(sq,L_y,d,+)$, and the number of eigenvalues with
different signs as $n_Z(sq,L_y,d,-)$. It is clear that 
\beq 
n_Z(sq,L_y,d,+) + n_Z(sq,L_y,d,-) = n_Z(sq,L_y,d) \ . 
\eeq
Define
\beq 
\Delta n_Z(sq,L_y,d) \equiv n_Z(sq,L_y,d,+) - n_Z(sq,L_y,d,-) 
\eeq
which gives
the number of partitions which have self-reflection symmetry. For example, among the partitions for the $L_y=2$ strip
in Fig. \ref{L2partitions}, the partitions $I$ and $12$ in ${\cal P}_{2,0}$, the third partition $\overline{12}$ in ${\cal P}_{2,1}$,
and the partition $\bar 1, \bar 2$ in ${\cal P}_{2,2}$ have
self-reflection symmetry. Among the partitions for the $L_y=3$
strip in Fig. \ref{L3partitions}, it includes the first partition $I$, the third partition $13$ and the fifth partition $123$
in ${\cal P}_{3,0}$, the second partition $\bar 2$, the sixth
partition $\overline{13}$ and the ninth partition $\overline{123}$ in ${\cal P}_{3,1}$, the second partition $\bar 1, \bar 3$ in ${\cal P}_{3,2}$, and the partition in ${\cal P}_{3,3}$. We list $\Delta n_Z(sq,L_y,d)$ for $1 \le L_y
\le 10$ in Table \ref{nzpmtable} in the appendix. The total number of these
partitions for each $L_y$ will be denoted as $\Delta N_{Z,L_y} = 2^{L_y}$. The
relations between $\Delta n_Z(sq,L_y,d)$ are
\beqs 
\Delta n_Z(sq,2n,0) & = & 2 \Delta n_Z(sq,2n-1,0) \ , \qquad
\mbox{for} \ 0<n \cr\cr 
\Delta n_Z(sq,2n,2m-1) & = & \Delta
n_Z(sq,2n,2m) \cr\cr & = & \Delta n_Z(sq,2n-1,2m-1) + \Delta n_Z(sq,2n-1,2m) \ , \qquad \mbox{for} \ 1 \le m \le n \cr\cr 
\Delta n_Z(sq,2n+1,2m) & = & \Delta
n_Z(sq,2n+1,2m+1) \cr\cr & = & \Delta n_Z(sq,2n,2m) + \Delta n_Z(sq,2n,2m+1) \ , \qquad \mbox{for} \ 0 \le m \le n \ .
\eeqs
We also list $n_Z(sq,L_y,d,+)$ and $n_Z(sq,L_y,d,-)$ for $2 \le L_y \le 10$ in
Table \ref{ntctable} in the appendix. Notice that
$n_Z(sq,L_y,0,+)$ is the number of $\lambda_{Z,sq,FF,L_y}$ proved in
\cite{ts}. Recall the numbers of $\lambda_{Z,sq,L_y,j}$ for the M\"obius strips
of the square lattice with coefficients $\pm c^{(d)}$, defined as
$n_{Z,Mb}(L_y,d,\pm) \equiv n_{Z,Mb}(sq,L_y,d,\pm)$, have been given in
\cite{cf}. With the eqs.  (\ref{cd0tran}) to (\ref{cdoddtran}), the relations
between $n_Z(sq,L_y,d,\pm)$ and $n_{Z,Mb}(sq,L_y,d,\pm)$ are
\beqs 
n_{Z,Mb}(sq,L_y,0,\pm) & = & n_Z(sq,L_y,0,\pm) + n_Z(sq,L_y,2,\mp) \cr\cr
n_{Z,Mb}(sq,L_y,k,\pm) & = & n_Z(sq,L_y,2k-1,\pm) + n_Z(sq,L_y,2k+2,\mp) \ ,
\cr\cr
& & {\rm for} \quad 1 \le k \le \Bigl [ \frac{L_y+1}{2} \Bigr ]  \ . 
\eeqs

For a strip graph $G_s$ of a lattice $\Lambda$ with given boundary conditions,
following our earlier notation \cite{a} we denote the sum of coefficients
(generalized multiplicities) $c_{G_s}$ of the $\lambda_{Z,G_s,j}$ as
$C_{Z,G_s}$.  This sum is independent of the length $m$ of the strip. For a
cyclic strip graph this is given by \cite{cf}
\beq
C_{Z,\Lambda, cyc.,L_y}= \sum_{d=0}^{L_y} c^{(d)} n_Z(L_y,d) = q^{L_y} 
\quad {\rm for} \ \ \Lambda=sq,tri,hc \ . 
\label{czsumcyc}
\eeq
For the M\"obius strip of the square lattice or the honeycomb lattice with
$L_y$ even, the sign changes of the eigenvalues of $\tilde T_{Z,\Lambda,L_y,d}$
can be considered as the sign changes of the coefficients. For these cases, the
sum of coefficients is given by Theorem 8 of Ref. \cite{cf}:
\beq
C_{Z,(sq,hc),L_y,Mb} \equiv \sum_{j=1}^{N_{Z,L_y,\lambda}} c_{Z,L_y,Mb,j} =
\sum_{d=0}^{d_{max}} \Delta n_{Z,Mb}(L_y,d) c^{(d)} =\cases{ q^{L_y/2} & for
even $L_y$ \cr
q^{(L_y+1)/2} & for odd $L_y$ \cr }
\label{ctsummb}
\eeq
where
\beq
d_{max}= \cases{ \frac{L_y}{2} & for even $L_y$ \cr
\frac{(L_y+1)}{2} & for odd $L_y$ \cr }
\label{dmax}
\eeq

In previous work we have given results for the determinants for various strip
graphs $G_s$ (e.g., \cite{s3a}) 
\beq
\det T_Z(G_s) = \prod_{j=1}^{N_{Z,G_s,\lambda}}
(\lambda_{Z,G_s,j})^{c_{Z,G_s,j}}
\label{detformz}
\eeq
and
\beq
\det T_P(G_s) = \prod_{j=1}^{N_{P,G_s,\lambda}}
(\lambda_{P,G_s,j})^{c_{P,G_s,j}} \ .
\label{detformp}
\eeq
and we shall extend these results to arbitrary width here.  In the present
context, these can be written, for cyclic strips, as
\beq
det(T_{Z,\Lambda,L_y}) = \prod_{d=0}^{L_y} [det(T_{Z,\Lambda,L_y,d})]^{c^{(d)}}
\label{detformzcyc}
\eeq
\beq
det(T_{T,\Lambda,L_y}) = \prod_{d=0}^{L_y} [det(T_{T,\Lambda,L_y,d})]^{c^{(d)}}
\ . 
\label{detformtcyc}
\eeq

 From eq. (\ref{zt}) it follows that \cite{a,hca}
\beq
\lambda_{Z,\Lambda,L_y,d,j} = v^{p L_y} \lambda_{T,\Lambda,L_y,d,j}
\label{lamzt}
\eeq
and
\beq
T_{Z,\Lambda,L_y,d} = v^{p L_y} T_{T,\Lambda,L_y,d}
\label{tztt}
\eeq
where
\beq
p = \cases{ 1 & if $\Lambda=sq$ \ or \ $tri$ \ or $G_D$ \cr
            2 & if $\Lambda=hc$ \cr }
\label{powerp}
\eeq
Note that the factor of $(x-1)$ in eq. (\ref{zt}) cancels the factor $1/(x-1)$
in eq. (\ref{tgsum_transfer}). 

\section{Properties of Transfer Matrices at Special Values of Parameters}

In this section we derive some properties of the transfer matrices 
$T_{Z,\Lambda,L_y,d}$ at special values of $q$ and $v$, and, correspondingly,
$T_{T,\Lambda,L_y,d}$ at special values of $x$ and $y$.  

\subsection{ $v=0$ }

 From (\ref{zfun}) or (\ref{cluster}) it follows that for any graph $G$, the
Potts model partition function $Z(G,q,v)$ satisfies 
\beq
Z(G,q,0)=q^{n(G)} \ . 
\label{zv0}
\eeq
Since this holds for arbitrary values of $q$, in the context of the lattice 
strips considered here, it implies 
\beq
(T_{Z,\Lambda,L_y,d})_{v=0} = 0 \quad {\rm for} \quad 1 \le d \le L_y
\label{tzlydv0}
\eeq
i.e. these are zero matrices. Secondly, restricting to cyclic strips for
simplicity, and using the basic results $n=L_yL_x=L_y m$ for $\Lambda=sq,tri$
and $n=2L_y m$ for $\Lambda=hc$, eq. (\ref{zv0}) implies that
\beq
Tr[(T_{Z,\Lambda,cyc.,L_y,q,v})^m]_{v=0} 
=\cases{q^{L_y m}&\ for \ $\Lambda=sq,tri$ \cr
             q^{2L_y m} &\ for \ $\Lambda=hc$  }
\label{TThc1xl} 
\eeq
With our explicit calculations, we find that
\beq
(T_{Z,\Lambda,L_y,0})_{jk} = 0 \quad {\rm for} \ \ v=0, \quad {\rm and} \quad
j \ge 2
\label{Tjkv0}
\eeq
i.e., all rows of these matrices except the first vanish, and
\beq
(T_{Z,\Lambda,L_y,0})_{11}=q^{p L_y} \quad {\rm for} \quad v=0 
\label{T11v0}
\eeq
where $p$ was given in eq. (\ref{powerp}). 
For this $v=0$ case, since all of the rows except the first are zero, the
elements $(T_{Z,\Lambda,L_y,0})_{1k}$ for $k \ge 2$ do not enter into
$Tr[(T_{Z,\Lambda,L_y,0})^m]$, which just reduces to the $m$'th power of the
(1,1) element:
\beq
Tr[(T_{Z,\Lambda,L_y,0})^m]=[(T_{Z,\Lambda,L_y,0})_{11}]^m \ . 
\label{tracev0}
\eeq
Corresponding to this, all of the eigenvalues $\lambda_{Z,\Lambda,L_y,d,j}$
vanish except for one, which is equal to $(T_{Z,\Lambda,L_y,0})_{11}$ in
eq. (\ref{T11v0}).  As will be seen, this is reflected in the property that
$det(T_{Z,\Lambda,L_y,d})$ has a nonzero power of $v$ as a factor for $L_y \ge
2$ for all of the lattice-$\Lambda$ strips considered here. The restriction
$L_y \ge 2$ is made because the strips of the triangular and honeycomb lattice
are well-defined without degenerating for $L_y \ge 2$ and, in the case of the
square lattice, for the case $L_y=1$, both of the transfer matrices reduce to
scalars, $T_{Z,sq,1,0}=q+v$ and $T_{Z,sq,1,1}=v$, the former of which is, in
general, nonzero at $v=0$.  Note that the condition $v=0$ is equivalent to 
the Tutte variable condition $y=1$. 

\subsection{ $q=0$ } 

Another fundamental relation that follows, e.g., by setting $q=0$ in eq. 
(\ref{cluster}), is 
\beq
Z(G,0,v)=0 \ . 
\label{zq0}
\eeq
Now the coefficients $c^{(d)}$ evaluated at $q=0$ satisfy \cite{cf}
\beq
c^{(d)}=(-1)^d \quad {\rm for} \quad q=0 \ .
\label{cdq0}
\eeq
Hence, in terms of transfer matrices, we derive the sum rule 
\beq
\sum_{0 \le d \le L_y, \ d \ even}  Tr[(T_{Z,\Lambda,L_y,d})^m] - 
\sum_{1 \le d \le L_y, \ d \ odd}   Tr[(T_{Z,\Lambda,L_y,d})^m] = 0 
\quad {\rm for} \ \ q=0 \ . 
\label{zq0tran}
\eeq
This is similar to a sum rule that we obtained in Ref. \cite{s5}.  Since it
applies for arbitrary $m$, it implies that there must be a pairwise
cancellation between various eigenvalues in different degree-$d$ subspaces, 
which, in turn, implies that at $q=0$ there are equalities between these
eigenvalues.  For example, for $L_y=1$ and $q=0$, $\lambda_{Z,sq,1,0}=q+v$ 
becomes equal to $\lambda_{Z,sq,1,1}=v$.  For the $L_y=2$ strips of all of the 
$\Lambda=sq,tri,hc$ lattices, $q=0$, two of the eigenvalues of 
$T_{Z,\Lambda,2,1}$ become equal to the eigenvalues of $T_{Z,\Lambda,2,0}$, 
while the third eigenvalue of $T_{Z,\Lambda,2,1}$ becomes equal to 
$\lambda_{Z,\Lambda,2,2}=v^{2p}$, and so forth for higher $L_y$.  Note that 
setting $q=0$ does not, in general, lead to any vanishing eigenvalues for
the $T_{Z,\Lambda,L_y,d}$ and hence our formulas below for 
$T_{Z,\Lambda,L_y,d}$ do not contain overall factors of $q$.  

In terms of Tutte polynomial parameters, the value $q=0$ implies $x=1$ (unless
$v=0$).  In contrast to the vanishing of $Z(G,q,v)$ at $q=0$, the Tutte
polynomial $T(G,1,y)$ is nonzero for general $y$.  This different behavior can
be traced to the feature that in eq. (\ref{zt}), $Z(G,q,v)$ is proportional to
$T(G,x,y)$ multiplied by the factor $(x-1)^{k(G)}=(x-1)$, so at $x=1$,
$Z(G,0,v)=0$ even if $T(G,1,y) \ne 0$.

\subsection{$q=1$} 

Evaluating eq. (\ref{zfun}) for $q=1$, one sees that the Kronecker delta
functions $\delta_{\sigma_i \sigma_j}=1$ for all pairs of adjacent vertices
$\langle i,j \rangle$; consequently, 
\beq
Z(G,1,v)=e^{e(G)K} = (1+v)^{e(G)} \ . 
\label{zq1}
\eeq
Now \cite{cf}
\beq
{\rm If} \quad q=1 \quad {\rm then} \quad
c^{(d)}= \cases{ 1 & if $d=0$ \ mod \ 3 \cr
                 0 & if $d=1$ \ mod \ 3 \cr
                -1 & if $d=2$ \ mod \ 3  }
\label{cdq1}
\eeq
Hence, in terms of transfer matrices, we derive the sum rule for the 
present cyclic lattice strips $G=\Lambda, L_y \times L_x,cyc.$ 
\beq
\sum_{0 \le d \le L_y, d=0 \ mod \ 3} Tr[(T_{Z,\Lambda,L_y,d})^m] - 
\sum_{2 \le d \le L_y, d=2 \ mod \ 3} Tr[(T_{Z,\Lambda,L_y,d})^m]  =
(1+v)^{e(G)} \quad {\rm for} \ \ q=1 \ , 
\label{zq1tran}
\eeq
Here the number of edges $e(G)$ for each type of cyclic strip is
\beq
e(G) = \cases{ (2L_y-1)m & if $\Lambda=sq$ \cr
            (3L_y-2)m & if $\Lambda=tri$ \cr
            (3L_y-1)m & if $\Lambda=hc$  \cr
            2L_ym      & if $\Lambda=G_D$ }
\label{edges}
\eeq
where $m$ is given in terms of $L_x$ by eq. (\ref{lxm}).  As before, the sum
rule (\ref{zq1tran}) implies relations between the eigenvalues of the various
transfer matrices $T_{Z,\Lambda,L_y,d}$. 

\subsection{$v=-1$, i.e., $y=0$} 

As discussed above, the special value $v=-1$, i.e., $y=0$, corresponds to the
zero-temperature Potts antiferromagnet, and in this case the Potts model
partition function reduces to the chromatic polynomial as indicated in
eq. (\ref{zp}).  In Refs. \cite{cf,dg,hca} we have determined how the
dimensions $n_Z(L_y,d)$ for the square, triangular, and honeycomb lattice
strips and $n_Z(G_D,L_y,d)$ for the self-dual strip of the square-lattice
reduce to dimensions $n_P(L_y,d)$ and $n_P(G_D,L_y,d)$, respectively.  In all
cases except for the lowest width $L_y=1$ for the square lattice strip, in each
degree-$d$ subspace for $0 \le d \le L_y-1$ for $\Lambda=sq,tri,hc$ and $1 \le
d \le L_y$ for $\Lambda=G_D$, some eigenvalues vanish so that
$n_P(\Lambda,L_y,d) < n_Z(\Lambda,L_y,d)$ and hence
$det(T_{Z,\Lambda,L_y,d})=0$.  This property is reflected in the powers of
$(v+1)$ and $y$ that appear, respectively, in our formulas below for
$det(T_{Z,\Lambda,L_y,d})$ and $det(T_{T,\Lambda,L_y,d})$.

\subsection{$q=-v$, i.e., $x=0$}

For the graph $G=G(V,E)$, setting $x=0$ and $y=1-q$ in the Tutte polynomial
$T(G,x,y)$ yields the flow polynomial $F(G,q)$, which counts the number of
nowhere-0 $q$-flows (without sinks or sources) that there are on $G$
\cite{boll}:
\beq
F(G,q)=(-1)^{e(G)-n(G)+1}T(G,0,1-q) \ . 
\label{ft}
\eeq
In Ref. \cite{f} we showed that the flow polynomial for cyclic lattice strips
has the form 
\beq
F(\Lambda,L_y \times L_x, cyc.,q) = \sum_{d=0}^{L_y}
c^{(d)} \sum_{j=1}^{n_F(\Lambda,L_y,d)}(\lambda_{F,\Lambda,L_y,d,j})^m
\label{fgsumcyc}
\eeq
or equivalently, 
\beq
F(\Lambda,L_y \times L_x, cyc.,q) = \sum_{d=0}^{L_y}
c^{(d)} \sum_{j=1}^{n_F(\Lambda,L_y,d)}Tr[(T_{F,\Lambda,L_y,d,j})^m] \ .
\label{fgsumcyctransfer}
\eeq
We determined the dimensions $n_F(\Lambda,L_y,d)$ and found that for all $L_y
\ge 1$, $n_F(\Lambda,L_y,d) < n_Z(\Lambda,L_y,d)$.  Thus, when one sets $x=0$
in the Tutte polynomial, or equivalently, $q=-v$ in the Potts model partition
function, some of the eigenvalues in each degree-$d$ subspace vanish, and hence
$det(T_{Z,\Lambda,L_y,d})$ and $det(T_{T,\Lambda,L_y,d})$ vanish.  This is
reflected in the powers of $(q+v)$ and $x$ that appear, respectively, in our 
formulas below for $det(T_{Z,\Lambda,L_y,d})$ and  $det(T_{T,\Lambda,L_y,d})$.
In passing, we note that for planar graphs $G_{pl}$, because of the duality 
relation (\ref{tdual}), the chromatic polynomial on $G$ is closely related to
the flow polynomial on $G^*$: $P(G,q)=qF(G^*,q)$. 

These results concerning evaluations of the Potts model partition function on
these lattice strips for special values of $q,v$, equivalent to evaluations of
the Tutte polynomial for special values of $x$ and $y$, help to explain the
types of factors that appear in the simple formulas that we obtain below for the
determinants $det(T_{Z,\Lambda,L_y,d})$ and $det(T_{T,\Lambda,L_y,d})$.

\section{General Results for Cyclic Strips of the Square Lattice}

In this section and the subsequent ones we present general results that we have
obtained, valid for arbitrarily large strip width $L_y$ (as well as arbitrarily
great length) for transfer matrices and their properties.  We begin with the
strip of the square lattice. 

\subsection{Determinants}

We find
\beq
det(T_{T,sq,L_y,d}) = (x^{L_y}y^{L_y-1})^{n_Z(L_y-1,d)}
\label{detTTsqld}
\eeq
where $n_Z(L_y,d)$ was given in eq. (\ref{nzlyd}). This applies for all $d$,
i.e., $0 \le d \le L_y$ since \cite{cf} $n_Z(L_y-1,d)=0$ for $d > L_y-1$.
This is equivalent to the somewhat more complicated expression for 
$det(T_{Z,sq,L_y,d})$:
\beq
det(T_{Z,sq,L_y,d}) = 
(v^{L_y})^{n_Z(L_y,d)} \biggl [ \Bigl ( 1+\frac{q}{v} \Bigr )^{L_y}
(1+v)^{L_y-1} \biggr ]^{n_Z(L_y-1,d)} \ . 
\label{detTZsqld}
\eeq
These determinant formulas can be explained as follows. By the arrangement of
the partitions as shown in Figs. \ref{L2partitions} and \ref{L3partitions}, the
matrix $J_{L_y,d,i,i+1}$ has the lower triangular form and the matrix
$D_{L_y,d,i}$ has the upper triangular form. From the definition of the
transfer matrix in eq.  (\ref{transfermatrix}), the determinant of
$T_{Z,sq,L_y,d}$ is the product of the diagonal elements of $H_{Z,sq,L_y,d}$
and $V_{Z,sq,L_y,d}$. The diagonal elements of $H_{Z,sq,L_y,d}$ has the form
$(1+v)^r$, where $r$ is the number of edges in the corresponding partition, and
the diagonal elements of $V_{Z,sq,L_y,d}$ has the form $v^{L_y} (1+q/v)^s$,
where $s$ is the number of vertices which does not connect to any other vertex
in the corresponding partition. Therefore, the power of $(1+v)$ in
eq. (\ref{detTZsqld}) is the sum of the number of edges of all the
$(L_y,d)$-partitions. Let us compare the $(L_y,d)$-partitions and $(L_y-1,d)$
partitions. Since each edge of a $(L_y,d)$-partition corresponds to converting
a vertex of a $(L_y-1,d)$-partition into a edge, the power of $(1+v)$ is
$(L_y-1)n_Z(L_y-1,d)$. It is clear that the power of $v$ is $L_y
n_Z(L_y,d)$. The power of $(1+q/v)$ in eq.  (\ref{detTZsqld}) is the sum of the
number of unconnected vertices of all the $(L_y,d)$-partitions. Compare the
$(L_y,d)$-partitions and $(L_y-1,d)$ partitions again. Since each unconnected
vertex of a $(L_y,d)$-partition corresponds to adding an unconnected vertex to
a $(L_y-1,d)$-partition with $L_y$ possible ways (between every two adjacent
vertices, above the highest vertex, and below the lowest vertex), the power of
$(1+q/v)$ is $L_y n_Z(L_y-1,d)$.

Next, taking into account that the generalized multiplicity of each
$\lambda_{Z,\Lambda,L_y,d}$ is $c^{(d)}$, we have, for the total determinant
\beqs
det(T_{Z,sq,L_y}) &=& \prod_{d=0}^{L_y} [det(T_{Z,sq,L_y,d})]^{c^{(d)}} \cr\cr
& = & \prod_{d=0}^{L_y}
    [(y-1)^{L_y}]^{n_Z(L_y,d)c^{(d)}} 
    [x^{L_y} y^{L_y-1}]^{n_Z(L_y-1,d)c^{(d)}} \cr\cr
& = & [(y-1)^{L_y}]^{\sum_{d=0}^{L_y} n_Z(L_y,d)c^{(d)}}
  [x^{L_y} y^{L_y-1}]^{\sum_{d=0}^{L_y} n_Z(L_y-1,d)c^{(d)}} \ . 
\label{detTZsqLaux}
\eeqs
Using eq. (\ref{czsumcyc}) together with eq. (4.2) from Ref. \cite{cf}, viz.,
$n_Z(L_y,d) = 0$ for $d > L_y$, so that
$\sum_{d=0}^{L_y} n_Z(L_y-1,d)c^{(d)}=\sum_{d=0}^{L_y-1} n_Z(L_y-1,d)c^{(d)}$,
we have, finally,
\beq
det(T_{Z,sq,L_y}) = [(y-1)^{L_y}]^{q^{L_y}} [x^{L_y} y^{L_y-1}]^{q^{L_y-1}} \ .
\label{detTZsqL}
\eeq
This agrees with the conjecture given as eq. (2.42) of our earlier Ref.
\cite{s3a} for the determinant of the transfer matrix of the cyclic strip of
the square lattice with arbitrary width $L_y$.

\subsection{Traces}

\begin{figure}
\unitlength 1mm \hspace*{5mm}
\begin{picture}(50,30)
\multiput(10,0)(0,6){2}{\circle{1}}
\multiput(10,18)(0,6){3}{\circle{1}}
\multiput(16,0)(0,6){2}{\circle{1}}
\multiput(16,18)(0,6){3}{\circle{1}}
\multiput(10,0)(0,6){2}{\line(1,0){6}}
\multiput(10,18)(0,6){3}{\line(1,0){6}}
\multiput(10,0)(6,0){2}{\line(0,1){6}}
\multiput(10,18)(6,0){2}{\line(0,1){12}}
\multiput(10,12)(6,0){2}{\makebox(0,0){{\small $\vdots$}}}
\multiput(4,0)(18,0){2}{\makebox(0,0){{\footnotesize $L_y$}}}
\multiput(4,6)(18,0){2}{\makebox(0,0){{\footnotesize $L_y-1$}}}
\multiput(4,18)(18,0){2}{\makebox(0,0){{\footnotesize $3$}}}
\multiput(4,24)(18,0){2}{\makebox(0,0){{\footnotesize $2$}}}
\multiput(4,30)(18,0){2}{\makebox(0,0){{\footnotesize $1$}}}
\put(36,15){\makebox(0,0){{\small $=$}}}
\multiput(56,0)(0,6){2}{\circle{1}}
\multiput(56,18)(0,6){3}{\circle{1}} \put(56,0){\line(0,1){6}}
\put(56,18){\line(0,1){12}} \put(56,12){\makebox(0,0){{\small
$\vdots$}}} \multiput(59,0)(0,6){2}{\oval(6,2)}
\multiput(59,18)(0,6){3}{\oval(6,2)}
\put(50,0){\makebox(0,0){{\footnotesize $L_y$}}}
\put(50,6){\makebox(0,0){{\footnotesize $L_y-1$}}}
\put(50,18){\makebox(0,0){{\footnotesize $3$}}}
\put(50,24){\makebox(0,0){{\footnotesize $2$}}}
\put(50,30){\makebox(0,0){{\footnotesize $1$}}}
\end{picture}

\caption{\footnotesize{$m=1$ graph for the cyclic square
lattice.}} \label{sqtrace}
\end{figure}
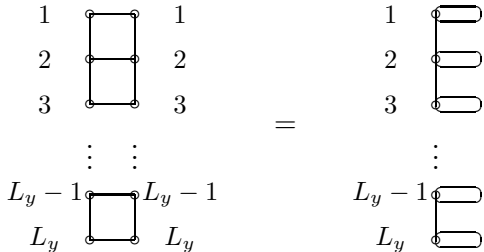

For the trace of the total transfer matrix, taking account of the fact that
each of the $\lambda_{X,\Lambda,d,j}$, $X=Z,T$, has multiplicity $c^{(d)}$, we
have
\beqs
Tr(T_{T,sq,L_y}) & = & \frac{1}{x-1} \sum_{d=0}^{L_y} c^{(d)} 
 \sum_{j=1}^{n_Z(L_y,d)}\lambda_{T,sq,L_y,d,j} 
= \frac{1}{x-1} \sum_{d=0}^{L_y} c^{(d)} Tr(T_{T,sq,L_y,d}) \cr\cr
& = & x^{L_y-1} y^{L_y} \ . 
\label{traceTTsq}
\eeqs
Note that $\sum_{d=0}^{L_y} c^{(d)} Tr(T_{T,sq,L_y,d})$ contains a factor of
$(x-1)$ which cancels the prefactor $1/(x-1)$.  Equivalently, we find 
\beqs
Tr(T_{Z,sq,L_y}) & = & \sum_{d=0}^{L_y} c^{(d)}\sum_{j=1}^{n_Z(L_y,d)} 
\lambda_{Z,sq,L_y,d,j}
=  \sum_{d=0}^{L_y} c^{(d)} Tr(T_{Z,sq,L_y,d}) \cr\cr
& = & q(v+q)^{L_y-1}(1+v)^{L_y} \ . 
\label{traceTZsq}
\eeqs
We derive this as follows.  The trace here is the $m=1$ case in eq.
(\ref{tgsum_transfer}) or (\ref{zgsum_transfer}), which corresponds to a
$L_y$-vertex tree with a loop attached to each vertex, as shown in
Fig. \ref{sqtrace}. The corresponding Tutte polynomial is $x^{L_y-1}y^{L_y}$.
In principle, we can also consider the $m=1$ case in
eq. (\ref{zgsum_transfermb}) or (\ref{tgsum_transfermb}) for the M\"obius
strips, but the result is not as simple as that listed here.

\subsection{Eigenvalue for $d=L_y$ for $\Lambda=sq,tri,hc$}

It was shown earlier \cite{a,tor} that the $\lambda$'s are the same for a given
lattice strip with cyclic, as compared with M\"obius, boundary conditions (and
it was shown that the $\lambda$'s for a strip with Klein bottle boundary
conditions are a subset of the $\lambda$'s for the same strip with torus
boundary conditions).  From eq. (\ref{nzlyd}) one knows that there is only one
$\lambda$ for degree $d=L_y$, which we denote $\lambda_{Z,\Lambda,L_y,L_y}$.
That is, for this value of $d$, the transfer matrix reduces to $1 \times 1$,
i.e. a scalar. We found that for a cyclic or M\"obius strip of the square,
triangular, or honeycomb lattice with width $L_y$,
\beq
\lambda_{T,\Lambda,L_y,L_y}=1 \ . 
\label{lamtutdly}
\eeq
(Indeed, we have found that this holds more generally for other strips of
regular lattices (e.g., \cite{ka}), but we shall not pursue this here.)  Note the
universality; i.e., this is the same for all of these lattices. In terms of
Potts model variables the results do not maintain the full universality:
\beq
\lambda_{Z,\Lambda,L_y,L_y}=v^{L_y} \quad {\rm for} \quad \Lambda=sq,tri
\label{lamdlysqtri}
\eeq
\beq
\lambda_{Z,\Lambda,L_y,L_y}=v^{2L_y} \quad {\rm for} \quad \Lambda=hc \ . 
\label{lamdlyhc}
\eeq

\subsection{Transfer Matrix for $d=L_y-1$, $\Lambda=sq$}

 From eq. (\ref{nzlyd}) it follows that for $\Lambda=sq,tri$ or $hc$, the
number of $\lambda_{X,L_y,d,j}$, $X=Z$ or $X=T$, for $d=L_y-1$ is
$n_Z(L_y,L_y-1)=2L_y-1$, i.e. the transfer matrix in this subspace,
$T_{X,\Lambda,d}$, $X=Z$ or $X=T$, is a (square) $(2L_y-1)$-dimensional matrix.
We recall that for $L_y=1$, $T_{T,sq,1,0}=x$.  For $L_y \ge 2$ we find the
following general formula, which we write for the Tutte polynomial,
since it has a somewhat simpler form.
\beq
(T_{T,sq,L_y,L_y-1})_{j,j}=1+x \quad {\rm for} \quad j = 1 \ \ {\rm and} \ \
j=L_y
\label{TTsqx1}
\eeq
\beq
(T_{T,sq,L_y,L_y-1})_{j,j}=2+x \quad {\rm for} \quad
L_y \ge 3 \ \ {\rm and} \ \ 2 \le j \le L_y-1
\label{TTsqx2}
\eeq
\beq
(T_{T,sq,L_y,L_y-1})_{j,j+1} = (T_{T,sq,L_y,L_y-1})_{j+1,j} = 1 
\quad {\rm for} \quad 1 \le j \le L_y-1
\label{TTsq1a}
\eeq
\beq
(T_{T,sq,L_y,L_y-1})_{j,j}=y \quad {\rm for} \quad L+1 \le j \le 2L_y-1
\label{TTsqy}
\eeq
\beq
(T_{T,sq,L_y,L_y-1})_{j,L_y+j}=(T_{T,sq,L_y,L_y-1})_{j+1,L_y+j}=
\frac{y}{y-1} \quad {\rm for} \quad 1 \le j \le L_y-1
\label{TTsqwa}
\eeq
\beq
(T_{T,sq,L_y,L_y-1})_{L_y+j,j}=(T_{T,sq,L_y,L_y-1})_{L_y+j,j+1}= y-1 
\quad {\rm for} \quad 1 \le j \le L_y-1
\label{TTsqva}
\eeq
with all other elements equal to zero.  The elements of $T_{Z,sq,L_y-1}$
are given by the relation (\ref{tztt}).  Thus, $T_{X,sq,L_y,L_y-1}$ for $X=T,Z$
consists of four submatrices:

\begin{enumerate}

\item

an upper left square submatrix with indices
$i,j$ in the ranges $1 \le i,j \le L_y$ and nonzero elements given by
eqs. (\ref{TTsqx1})-(\ref{TTsq1a}) 

\item 

a lower right square submatrix with
indices in the ranges $L_y+1 \le i,j \le 2L_y-1$ and nonzero elements given by
eqs. (\ref{TTsqy})

\item 

an upper right rectangular submatrix with nonzero
elements given by eq. (\ref{TTsqwa})

\item 

a lower left
rectangular submatrix with nonzero elements given by eq. (\ref{TTsqva}). 

\end{enumerate} 

For general $x$ and $y$, $T_{T,sq,L_y,L_y-1}$ has rank equal to its dimension,
$2L_y-1$.  In the special case $y=0$ which yields the chromatic polynomial, the
rank is reduced to $L_y$ for general $x$ (and may be reduced further for
particular $x$).  In the special case $x=0$ which yields the flow polynomial,
the rank is reduced to $L_y-1$ (and may be reduced further for particular
$y$).

We illustrate these general formulas for the cases $L_y=2,3,4$.  For this
purpose we introduce the abbreviations
\beq
x_1=1+x, \quad x_2=2+x, \quad r_d=\frac{y}{y-1}
\label{abbrev1}
\eeq
and use $v=y-1$ as before.
\beq
T_{T,sq,2,1} = \left( \begin{array}{ccc}
    x_1 & 1   &  r_d \\
    1   & x_1 &  r_d \\
    v   & v   &  y \end{array} \right )
\label{TTsq21}
\eeq
\beq
T_{T,sq,3,2} = \left( \begin{array}{ccccc}
    x_1  & 1   &  0  & r_d  & 0 \\
    1    & x_2 &  1  & r_d  & r_d \\
    0    & 1   & x_1 & 0    & r_d \\
    v    & v   & 0   & y    & 0 \\
    0    & v   & v   & 0    & y \end{array} \right )
\label{TTsq32}
\eeq
\beq
T_{T,sq,4,3} = \left( \begin{array}{ccccccc}
    x_1 & 1  &  0 & 0  & r_d & 0  & 0  \\
    1  & x_2 &  1 & 0  & r_d & r_d & 0  \\
    0  & 1  & x_2 & 1  & 0  & r_d & r_d \\
    0  & 0  & 1  & x_1 & 0  & 0  & r_d \\
    v  & v  & 0  & 0  & y  & 0  & 0  \\
    0  & v  & v  & 0  & 0  & y  & 0  \\
    0  & 0  & v  & v  & 0  & 0  & y \end{array} \right )
\label{TTsq43}
\eeq
Thus, in general, the upper left-hand submatrix has a main diagonal with end
entries equal to $x_1$ and interior entries equal to $x_2$.  Adjacent to this
main diagonal are two diagonals whose entries are 1, and the rest of the
submatrix is comprised of triangular regions filled with 0's.  The upper
right-hand submatrix has a band of two diagonals whose entries are $r_d$
together with triangular regions filled with 0's.  The lower left-hand
submatrix has a band of two diagonals whose entries are $v$ together with
triangular regions filled with 0's.  And finally, in the right-hand lower
submatrix the entries on the main diagonal are equal to $y$ and the rest of
this submatrix is made up of triangular regions of 0's.  For the lowest values
$L_y=1,2$, some of these parts, such as the triangular regions of zeros, are
not present.

Using eq. (\ref{tztt}), it is straightforward to obtain the
$T_{Z,sq,L_y,L_y-1}$ from these $T_{T,sq,L_y,L_y-1}$ matrices.  For example, 
\beq
T_{Z,sq,2,1}= \left( \begin{array}{ccc}
  v(q+2v)   & v^2      & v(1+v)    \\
   v^2      & v(q+2v)  & v(1+v)    \\
   v^3      & v^3      & v^2(1+v)  \end{array} \right )
\label{TZsq21}
\eeq

Having determined the general form of $T_{T,sq,L_y,d}$ for $d=L_y-1$, we
calculate its eigenvalues $\lambda_{T,sq,L_y,L_y-1,j}$.  We find that these
consist of one that is the same independent of $L_y$, namely
\beq
\lambda_{T,sq,L_y,L_y-1,1} = x
\label{lamtutsqdm1x}
\eeq
together with $2(L_y-1)$ quadratic roots.  It is convenient to label these as
($j,\pm$) with $2 \le j \le L_y$.  We find
\beq
\lambda_{T,sq,L_y,L_y-1,j,\pm} = \frac{1}{2}\biggl [ x + y + b_{sq,L_y,j} \pm
\sqrt{(x + y + b_{sq,L_y,j})^2 - 4xy} \ \biggr ] \quad {\rm for} \quad
2 \le j \le L_y
\label{lamtutsqdm1j}
\eeq
where
\beq
b_{sq,L_y,j} \equiv a_{sq,L_y,j}-1 =
 4 \cos^2 \Bigl ( \frac{(L_y+1-j)\pi}{2L_y} \Bigr ) \ . 
\label{blyj}
\eeq
Note that the product
$(\lambda_{T,sq,L_y,L_y-1,j,+})(\lambda_{T,sq,L_y,L_y-1,j,-})$ is independent
of $j$:
\beq (\lambda_{T,sq,L_y,L_y-1,j,+})(\lambda_{T,sq,L_y,L_y-1,j,-}) = xy \ ,
\quad {\rm for} \quad 2 \le j \le L_y \ .
\label{lamtutsqprod}
\eeq
This generalizes to the Tutte polynomial our determination of the $L_y-1$ terms
$\lambda_{P,sq,L_y,L_y-1,j}$ for the chromatic polynomial in eqs. (7.1.2) and
(7.1.3) of Ref. \cite{s5}, and $a_{sq,L_y,j}$ was given in eq. (7.1.3) of that
paper.  Note that in the special case $y=0$ ($v=-1$) in which the Tutte
polynomial or Potts model partition function reduces to the chromatic
polynomial, one of each of the $L_y-1$ pairs of quadratic roots vanishes and
the other becomes $x+b_{sq,L_y,j}$ for the Tutte polynomial, or equivalently,
$(-1)^{L_y}(a_{sq,L_y,j}-q)$ for the Potts model, in agreement with eq. (7.1.2)
of Ref. \cite{s5}.  The term $(-1)^{L_y}(a_{sq,L_y,1}-q) = (-1)^{L_y}(1-q)$ in
eq. (7.1.2) of Ref. \cite{s5} corresponds to the $v=-1$ special case of
eq. (\ref{lamtutsqdm1x}).

As corollaries of our general result for $T_{T,sq,L_y,L_y-1}$ we calculate the
trace and determinant.  For this purpose, we note that
\beq
\sum_{j=2}^{L_y} b_{sq,L_y,j} = 4\sum_{j=2}^{L_y}
\cos^2 \Bigl ( \frac{(L_y+1-j)\pi}{2L_y} \Bigl ) = 2(L_y-1)
\label{bljsum}
\eeq
\beq
det(T_{T,sq,L_y,L_y-1}) = x^{L_y}y^{L_y-1}
\label{detTsqdm1}
\eeq
which is a special case of eq. (\ref{detTTsqld}), and 
\beq
Tr(T_{T,sq,L_y,L_y-1}) = x + (L_y-1)(2+x+y) = (L_y-1)(2+y)+L_y x \ . 
\label{traceTsqdm1}
\eeq
In terms of Potts model quantities these results are 
\beq
\lambda_{Z,sq,L_y,L_y-1,1}=v^{L_y-1}(v+q)
\label{lamsqdm1x}
\eeq
\beqs 
\lambda_{Z,sq,L_y,L_y-1,j\pm } & = & \frac{v^{L_y-1}}{2} \biggl [ q +
v(v+b_{sq,L_y,j}+2) \pm \sqrt{[q + v(v+b_{sq,L_y,j}+2)]^2 - 4v(v+q)(v+1)} \ 
\biggr ] \cr\cr & & \quad {\rm for} \quad 2 \le j \le L_y
\label{lamsqdm1j}
\eeqs
\beq
det(T_{Z,sq,L_y,L_y-1}) = v^{2L_y(L_y-1)} (v+q)^{L_y} (v+1)^{L_y-1} 
\label{detTZsqdm1}
\eeq
and
\beq
Tr(T_{Z,sq,L_y,L_y-1}) = v^{L_y-1}\biggl [ (4L_y-3)v +(L_y-1)v^2 + L_y q 
\biggr ] \ . 
\label{traceTZsqdm1}
\eeq

\section{General Results for Cyclic Strips of the Triangular Lattice}

\subsection{Determinants}

We find
\beq
det(T_{T,tri,L_y,d}) = (x^{L_y}y^{2(L_y-1)})^{n_Z(L_y-1,d)} \ . 
\label{detTTtrild}
\eeq
Equivalently,
\beq
det(T_{Z,tri,L_y,d}) = (v^{L_y})^{n_Z(L_y,d)}
\biggl [ \Bigl ( 1+\frac{q}{v} \Bigr )^{L_y} (1+v)^{2(L_y-1)} 
\biggr ]^{n_Z(L_y-1,d)}
\label{detTZtrild}
\eeq
By the same argument as before, the powers of $v$ and $(1+q/v)$ are equal to
the corresponding powers for the square lattice case.  Comparing
$V_{Z,sq,L_y,d}$ and $V_{Z,tri,L_y,d}$ in eq. (\ref{HVmatrix}), one sees that a
set of $(I+vJ_{L_y,d,i,i+1})$ has been included for the triangular lattice, so
that the power of $(1+v)$ becomes twice of the corresponding power for the
square lattice.

Taking into account that the multiplicity of each
$\lambda_{X,\Lambda,L_y,d,j}$, $X=Z,T$ is $c^{(d)}$, it follows that the
total determinant is 
\beq
det(T_{T,tri,L_y}) \equiv \prod_{d=0}^{L_y} [det(T_{T,tri,L_y,d})]^{c^{(d)}} =
(x^{L_y}y^{2(L_y-1)})^{q^{L_y-1}} \ . 
\label{detTTtri}
\eeq
Equivalently,
\beq
det(T_{Z,tri,L_y}) \equiv \prod_{d=0}^{L_y} [det(T_{Z,tri,L_y,d})]^{c^{(d)}} =
(v^{L_y})^{q^{L_y}} \biggl [ \Bigl ( 1+\frac{q}{v} \Bigr )^{L_y}
(1+v)^{2(L_y-1)} \biggr ]^{q^{L_y-1}} \ . 
\label{detTZsq}
\eeq

\subsection{Traces}

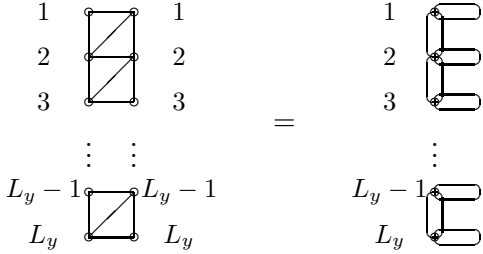
\begin{figure}
\unitlength 1mm \hspace*{5mm}
\begin{picture}(50,30)
\multiput(10,0)(0,6){2}{\circle{1}}
\multiput(10,18)(0,6){3}{\circle{1}}
\multiput(16,0)(0,6){2}{\circle{1}}
\multiput(16,18)(0,6){3}{\circle{1}}
\multiput(10,0)(0,6){2}{\line(1,0){6}}
\multiput(10,18)(0,6){3}{\line(1,0){6}}
\multiput(10,0)(6,0){2}{\line(0,1){6}}
\multiput(10,18)(6,0){2}{\line(0,1){12}} \put(10,0){\line(1,1){6}}
\multiput(10,18)(0,6){2}{\line(1,1){6}}
\multiput(10,12)(6,0){2}{\makebox(0,0){{\small $\vdots$}}}
\multiput(4,0)(18,0){2}{\makebox(0,0){{\footnotesize $L_y$}}}
\multiput(4,6)(18,0){2}{\makebox(0,0){{\footnotesize $L_y-1$}}}
\multiput(4,18)(18,0){2}{\makebox(0,0){{\footnotesize $3$}}}
\multiput(4,24)(18,0){2}{\makebox(0,0){{\footnotesize $2$}}}
\multiput(4,30)(18,0){2}{\makebox(0,0){{\footnotesize $1$}}}
\put(36,15){\makebox(0,0){{\small $=$}}}
\multiput(56,0)(0,6){2}{\circle{1}}
\multiput(56,18)(0,6){3}{\circle{1}} \put(56,3){\oval(2,6)}
\multiput(56,21)(0,6){2}{\oval(2,6)}
\put(56,12){\makebox(0,0){{\small $\vdots$}}}
\multiput(59,0)(0,6){2}{\oval(6,2)}
\multiput(59,18)(0,6){3}{\oval(6,2)}
\put(50,0){\makebox(0,0){{\footnotesize $L_y$}}}
\put(50,6){\makebox(0,0){{\footnotesize $L_y-1$}}}
\put(50,18){\makebox(0,0){{\footnotesize $3$}}}
\put(50,24){\makebox(0,0){{\footnotesize $2$}}}
\put(50,30){\makebox(0,0){{\footnotesize $1$}}}
\end{picture}

\caption{\footnotesize{$m=1$ graph for the cyclic triangular
lattice.}} \label{tritrace}
\end{figure}

For the total trace, taking account of the fact that each of the
$\lambda_{X,\Lambda,d,j}$ has multiplicity $c^{(d)}$, we have
\beqs
Tr(T_{T,tri,L_y}) & = & \frac{1}{x-1}\sum_{d=0}^{L_y} c^{(d)} 
\sum_{j=1}^{n_Z(L_y,d)} \lambda_{T,tri,L_y,d,j}  = 
\frac{1}{x-1}\sum_{d=0}^{L_y} c^{(d)} Tr(T_{T,tri,L_y,d}) \cr\cr
 & = & (x+y)^{L_y-1} y^{L_y} \ . 
\label{traceTTtri}
\eeqs
Equivalently,
\beqs
Tr(T_{Z,tri,L_y}) & = & \sum_{d=0}^{L_y}  c^{(d)} 
\sum_{j=1}^{n_Z(L_y,d)}\lambda_{Z,sq,L_y,d,j} =
\sum_{d=0}^{L_y} c^{(d)} Tr(T_{Z,tri,L_y,d}) \cr\cr
& = & q(q+2v+v^2)^{L_y-1}(v+1)^{L_y} \ . 
\label{traceTZtri}
\eeqs
Again the trace here is related to the $m=1$ case in eq. (\ref{tgsum_transfer})
or (\ref{zgsum_transfer}), which corresponds to a $L_y$-vertex tree with each
edge doubled and with a loop attached to each vertex as shown in
Fig. \ref{tritrace}. Therefore, the Tutte polynomial of this graph is
$(x+y)^{L_y-1}y^{L_y}$.

\subsection{Transfer Matrix for $d=L_y-1$, $\Lambda=tri$}

As before, from eq. (\ref{nzlyd}), we know that the dimension of this transfer
matrix is $2L_y-1$. The first nontrivial case is $L_y=2$. For $L_y \ge 2$ we
find the following general formula, which we write for the Tutte polynomial,
since it has a somewhat simpler form.
\beq
(T_{T,tri,L_y,L_y-1})_{1,1}=2+x+y
\label{TTtri_11}
\eeq
\beq
(T_{T,tri,L_y,L_y-1})_{j,j}=3+x+y \quad {\rm for} \quad 2 \le j \le L_y-1
\label{TTtri_lud}
\eeq
\beq
(T_{T,tri,L_y,L_y-1})_{L_y,L_y}=1+x
\label{TTtri_x1}
\eeq
\beq
(T_{T,tri,L_y,L_y-1})_{1+j,j}= 1 \quad {\rm for} \quad 1 \le j \le L_y-1
\label{TTtri_1}
\eeq
\beq
(T_{T,tri,L_y,L_y-1})_{j,k}= 3+x+2y \quad {\rm for} \quad 1 \le j \le L_y-2 \quad
{\rm and} \quad j+1 \le k \le L_y-1
\label{TTtri_lu}
\eeq
\beq
(T_{T,tri,L_y,L_y-1})_{j,L_y}=1+x+y \quad {\rm for} \quad 1 \le j \le L_y-1
\label{TTtri_cu}
\eeq
\beq
(T_{T,tri,L_y,L_y-1})_{1+j,L_y+j}=\frac{y}{y-1} \quad {\rm for} \quad 1 \le j
\le L_y-1
\label{TTtri_rd}
\eeq
\beq
(T_{T,tri,L_y,L_y-1})_{j,L_y+k}= \frac{y(1+y)}{y-1} \quad {\rm for} \quad
1 \le j \le L_y-1 \quad {\rm and} \quad j\le k \le L_y-1
\label{TTtri_ru}
\eeq
\beq
(T_{T,tri,L_y,L_y-1})_{j,k}=(y-1)(2+y) \quad {\rm for} \quad
1 \le j \le L_y-2 \quad {\rm and} \quad j+1 \le k \le L_y-1
\label{TTtri_ld}
\eeq
\beq
(T_{T,tri,L_y,L_y-1})_{L_y+j,1+k}=(y-1)(3+x+2y) \quad {\rm for}
\quad 1 \le j \le L_y-2 \quad {\rm and} \quad j \le k \le L_y-1
\label{TTtri_ll}
\eeq
\beq
(T_{T,tri,L_y,L_y-1})_{L_y,j}=(y-1)(1+x+y) \quad {\rm for} \quad 1 \le j \le L_y-1
\label{TTtri_cl}
\eeq
\beq
(T_{T,tri,L_y,L_y-1})_{j,L_y+k}=y(1+y) \quad {\rm for} \quad 1 \le j \le k \le L_y-1
\label{TTtri_rl}
\eeq
with all other elements equal to zero.  The elements of $T_{Z,tri,L_y,L_y-1}$
are given by the relation (\ref{tztt}).  Thus, $T_{X,tri,L_y,L_y-1}$ can again
be usefully viewed as consisting of various submatrices.  

For general $x$ and $y$, $T_{T,tri,L_y,L_y-1}$ has rank equal to its dimension,
$2L_y-1$.  In the special case $y=0$ (chromatic polynomial), the rank is
reduced to $L_y$ for general $x$ (and may be reduced further for particular
$x$).  In the special case $x=0$ (flow polynomial), the rank is reduced to
$2(L_y-1)$ (and may be reduced further for particular $y$).

We illustrate these general formulas with some explicit examples for
$L_y=2,3,4$.  For compactness of notation, we use the abbreviations
\beq
t_{11}=2+x+y, \quad l_{ud}=3+x+y, \quad \quad l_u=3+x+2y, 
\label{abbrev2}
\eeq
\beq
l_l=(y-1) l_u = (y-1)(3+x+2y), \quad l_d=(y-1)(y+2), 
\label{abbrev3}
\eeq
\beq
c_u=1+x+y, \quad c_l=(y-1) c_u = (y-1)(1+x+y), 
\label{abbrev4}
\eeq
\beq
r_u= \frac{y(y+1)}{y-1}, \quad r_l=y(y+1) 
\label{abbrev5}
\eeq
(with $x_1=1+x$ and $r_d=y/(y-1)$ as above).  Then 
\beq
T_{T,tri,2,1} = \left( \begin{array}{ccc}
    t_{11} & c_u & r_u \\
    1   & x_1 & r_d \\
    l_d  & c_l & r_l \end{array} \right )
\label{TTtri21}
\eeq
\beq
T_{T,tri,3,2} = \left( \begin{array}{ccccc}
   t_{11} & l_u & c_u & r_u & r_u \\
    1     & l_{ud}& c_u & r_d & r_u \\
    0     & 1  & x_1 & 0  & r_d \\
    l_d   & l_l & c_l & r_l & r_l \\
    0     & l_d & c_l & 0  & r_l \end{array} \right )
\label{TTtri32}
\eeq
\beq
T_{T,tri,4,3} = \left( \begin{array}{ccccccc}
    t_{11} & l_u & l_u & c_u & r_u & r_u & r_u \\
    1   & l_{ud}& l_u & c_u & r_d & r_u & r_u \\
    0   & 1  & l_{ud}& c_u & 0  & r_d & r_u \\
    0   & 0  & 1  & x_1 & 0  & 0  & r_d \\
    l_d & l_l & l_l & c_l & r_l & r_l & r_l \\
    0   & l_d & l_l & c_l & 0  & r_l & r_l \\
    0   & 0  & l_d & c_l & 0  & 0  & r_l \end{array} \right )
\label{TTtri43}
\eeq
It is straightforward to obtain the $T_{Z,tri,L_y,L_y-1}$ from these
$T_{T,tri,L_y,L_y-1}$ matrices; for example, 
\beq
T_{Z,tri,2,1}= \left( \begin{array}{ccc}
   v(q+4v+v^2)   & v(q+3v+v^2)   & v(2+v)(1+v)  \\
   v^2           & v(q+2v)       & v(1+v)  \\
   v^3(3+v)      & v^2(q+3v+v^2) & v^2(2+v)(1+v) \end{array} \right )
\label{TZtri21}
\eeq

\section{General Results for Cyclic Strips of the Honeycomb Lattice}

\subsection{Determinants}

We find
\beq 
det(T_{T,hc,L_y,d}) = (x^{2L_y}y^{L_y-1})^{n_Z(L_y-1,d)} \ . 
\label{detTThcld} 
\eeq
Equivalently,
\beq 
det(T_{Z,hc,L_y,d}) = (v^{2L_y})^{n_Z(L_y,d)} \biggl [ \Bigl (
1+\frac{q}{v} \Bigr )^{2L_y} (1+v)^{L_y-1} \biggr ]^{n_Z(L_y-1,d)} \ . 
\label{detTZhcld} 
\eeq
This can be understood as follows: by an argument similar to that given before,
the power of $(1+v)$ is the same as for the square lattice case. Comparing
$T_{Z,sq,L_y,d}$ and $T_{Z,hc,L_y,d}$ in eq. (\ref{transfermatrix}), one sees
that $V_{Z,hc,L_y,d} = V_{Z,sq,L_y,d}$ has been multiplied twice for the
honeycomb lattice, so that the powers of $v$ and $(1+q/v)$ become twice of the
corresponding powers for the square lattice. 

Taking into account that the multiplicity of each
$\lambda_{X,\Lambda,L_y,d,j}$, $X=Z,T$ is $c^{(d)}$, it follows that the total
determinant for the $hc$ lattice is
\beq det(T_{T,hc,L_y}) \equiv \prod_{d=0}^{L_y}
[det(T_{T,hc,L_y,d})]^{c^{(d)}} = (x^{2L_y}y^{L_y-1})^{q^{L_y-1}} \ . 
\label{detTThc} 
\eeq
Equivalently,
\beq 
det(T_{Z,hc,L_y}) \equiv \prod_{d=0}^{L_y}
[det(T_{Z,hc,L_y,d})]^{c^{(d)}} = (v^{2L_y})^{q^{L_y}} \biggl [ \Bigl (
1+\frac{q}{v} \Bigr )^{2L_y} (1+v)^{L_y-1} \biggr ]^{q^{L_y-1}} \ . 
\label{detTZhc} 
\eeq

Summarizing the connections between the determinants of the transfer matrices
for the three lattice strips, $det(T_{T,tri,L_y,d})$ is related to
$det(T_{T,sq,L_y,d})$ by the replacement $y \to y^2$ (holding $x$ fixed), while
$det(T_{T,hc,L_y,d})$ is related to $det(T_{T,sq,L_y,d})$ by the replacement $x
\to x^2$ (holding $y$ fixed).  This, together with the fact that $n_Z(L_y,d)$
is the same for all of these three lattices means that the total determinants
$det(T_{T,tri,L_y})$ and $det(T_{T,hc,L_y})$ are related to $det(T_{T,sq,L_y})$
by the same respective replacements.  Correspondingly, $det(T_{Z,tri,L_y,d})$
is related to $det(T_{Z,sq,L_y,d})$ by the replacement of $(1+v)$ by $(1+v)^2$
and $det(T_{Z,hc,L_y,d})$ is related to $det(T_{Z,sq,L_y,d})$ by the
replacements of the respective factors $v$ by $v^2$ (cf. eq. (\ref{powerp}))
and $(1+\frac{q}{v})$ by $(1 + \frac{q}{v})^2$.

\subsection{Traces}

\begin{figure}
\unitlength 1mm \hspace*{5mm}
\begin{picture}(70,36)
\multiput(10,0)(0,6){2}{\circle{1}}
\multiput(10,18)(0,6){4}{\circle{1}}
\multiput(16,0)(0,6){2}{\circle{1}}
\multiput(16,18)(0,6){4}{\circle{1}}
\multiput(22,0)(0,6){2}{\circle{1}}
\multiput(22,18)(0,6){4}{\circle{1}}
\multiput(10,0)(0,6){2}{\line(1,0){12}}
\multiput(10,18)(0,6){4}{\line(1,0){12}}
\multiput(10,0)(12,0){2}{\line(0,1){6}}
\multiput(10,18)(12,0){2}{\line(0,1){6}}
\multiput(10,30)(12,0){2}{\line(0,1){6}}
\put(16,24){\line(0,1){6}} \put(16,12){\makebox(0,0){{\small
$\vdots$}}} \multiput(4,0)(24,0){2}{\makebox(0,0){{\footnotesize
$L_y$}}} \multiput(4,6)(24,0){2}{\makebox(0,0){{\footnotesize
$L_y-1$}}} \multiput(4,18)(24,0){2}{\makebox(0,0){{\footnotesize
$4$}}} \multiput(4,24)(24,0){2}{\makebox(0,0){{\footnotesize
$3$}}} \multiput(4,30)(24,0){2}{\makebox(0,0){{\footnotesize
$2$}}} \multiput(4,36)(24,0){2}{\makebox(0,0){{\footnotesize
$1$}}} \put(42,18){\makebox(0,0){{\small $=$}}}
\multiput(62,0)(0,6){2}{\circle{1}}
\multiput(68,0)(0,6){2}{\circle{1}}
\multiput(62,18)(0,6){4}{\circle{1}}
\multiput(68,18)(0,6){4}{\circle{1}} \put(62,0){\line(0,1){6}}
\put(62,18){\line(0,1){6}} \put(68,24){\line(0,1){6}}
\put(62,30){\line(0,1){6}} \multiput(65,0)(0,6){2}{\oval(6,2)}
\multiput(65,18)(0,6){4}{\oval(6,2)}
\put(65,12){\makebox(0,0){{\small $\vdots$}}}
\put(56,0){\makebox(0,0){{\footnotesize $L_y$}}}
\put(56,6){\makebox(0,0){{\footnotesize $L_y-1$}}}
\put(56,18){\makebox(0,0){{\footnotesize $4$}}}
\put(56,24){\makebox(0,0){{\footnotesize $3$}}}
\put(56,30){\makebox(0,0){{\footnotesize $2$}}}
\put(56,36){\makebox(0,0){{\footnotesize $1$}}}
\end{picture}

\caption{\footnotesize{$m=1$ graph for the cyclic honeycomb
lattice with even $L_y$.}} \label{hctrace}
\end{figure}
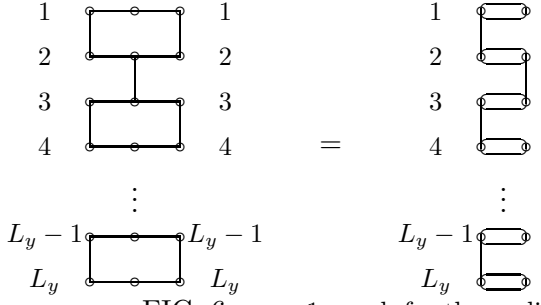

For the total trace, taking account of the fact that each of the
$\lambda_{X,\Lambda,d,j}$ has multiplicity $c^{(d)}$, we have
\beqs Tr(T_{T,hc,L_y}) & = & \frac{1}{x-1} \sum_{d=0}^{L_y} c^{(d)}
\sum_{j=1}^{n_Z(L_y,d)} \lambda_{T,hc,L_y,d,j} 
= \frac{1}{x-1} \sum_{d=0}^{L_y} c^{(d)} Tr(T_{T,hc,L_y,d}) \cr\cr
& = & x^{L_y-1} (x+y)^{L_y} \ . 
\label{traceTThc} 
\eeqs
Equivalently,
\beqs 
Tr(T_{Z,hc,L_y}) & = & \sum_{d=0}^{L_y} c^{(d)} \sum_{j=1}^{n_Z(L_y,d)}
\lambda_{Z,hc,L_y,d,j} = \sum_{d=0}^{L_y} c^{(d)} Tr(T_{Z,hc,L_y,d}) \cr\cr
& = & q(q+v)^{L_y-1}(q+2v+v^2)^{L_y} 
\label{traceTZhc} 
\eeqs
The trace here is related to the $m=1$ case in eq. (\ref{tgsum_transfer}) or
(\ref{zgsum_transfer}), which corresponds to a $2L_y$-vertex tree with each odd
edge doubled as shown in Fig. \ref{hctrace}. Therefore, the Tutte polynomial of
this graph is $x^{L_y-1}(x+y)^{L_y}$.

\subsection{Transfer Matrix for $d=L_y-1$, $\Lambda=hc$}

As before, from eq. (\ref{nzlyd}), we know that the dimension of
this transfer matrix is $2L_y-1$. The first nontrivial case is
$L_y=2$. Recall in eq. (\ref{transfermatrix}) the transfer matrix
for the honeycomb lattice, $T_{Z,hc,L_y,L_y-1}$, is the product of
$T_{Z,hc,L_y,L_y-1,1}$ and $T_{Z,hc,L_y,L_y-1,2}$. For $L_y \ge 2$ we find
the following general formula, which we write for the Tutte
polynomial.
\beq (T_{T,hc,L_y,L_y-1,1})_{j,j}=1+x \quad {\rm for} \quad 1 \le j
\le L_y-1 \label{TThc1x1} \eeq
\beq (T_{T,hc,L_y,L_y-1,1})_{L_y,L_y}= \cases{ 1+x & for $L_y$ even \cr
                                       x   & for $L_y$ odd  }
\label{TThc1xlb} \eeq
\beq (T_{T,hc,L_y,L_y-1,1})_{2j-1,2j}= (T_{T,hc,L_y,L_y-1,1})_{2j,2j-1}= 1
\quad {\rm for} \quad 1 \le j \le [L_y/2] \label{TThc1_1} \eeq
\beq (T_{T,hc,L_y,L_y-1,1})_{L_y+2j-1,L_y+2j-1}=y \quad {\rm for} \quad 1
\le j \le [L_y/2] \label{TThc1ya} \eeq
\beq (T_{T,hc,L_y,L_y-1,1})_{L_y+2j,L_y+2j}=1 \quad {\rm for} \quad 1 \le
j \le [(L_y-1)/2] \label{TThc1yb} \eeq
\beq
(T_{T,hc,L_y,L_y-1,1})_{2j-1,L_y+2j-1}=(T_{T,hc,L_y,L_y-1,1})_{2j,L_y+2j-1}
=\frac{y}{y-1} \quad {\rm for} \quad 1 \le j \le [L_y/2]
\label{TThc1wa} \eeq
\beq (T_{T,hc,L_y,L_y-1,1})_{2j,L_y+2j}=(T_{T,hc,L_y,L_y-1,1})_{2j+1,L_y+2j}
=\frac{1}{y-1} \quad {\rm for} \quad 1 \le j \le [(L_y-1)/2]
\label{TThc1wb} \eeq
\beq
(T_{T,hc,L_y,L_y-1,1})_{L_y+2j-1,2j-1}=(T_{T,hc,L_y,L_y-1,1})_{L_y+2j-1,2j}
=y-1 \quad {\rm for} \quad 1 \le j \le [L_y/2] \label{TThc1v} \eeq
\beq (T_{T,hc,L_y,L_y-1,2})_{1,1}=x \label{TThc2x} \eeq
\beq (T_{T,hc,L_y,L_y-1,2})_{j,j}=1+x \quad {\rm for} \quad 2 \le j
\le L_y-1 \label{TThc2x1} \eeq
\beq (T_{T,hc,L_y,L_y-1,2})_{L_y,L_y}= \cases{ 1+x & for $L_y$ odd \cr
                                       x   & for $L_y$ even  }
\label{TThc2xl} \eeq
\beq (T_{T,hc,L_y,L_y-1,2})_{2j,2j+1}= (T_{T,hc,L_y,L_y-1,1})_{2j+1,2j}= 1
\quad {\rm for} \quad 1 \le j \le [(L_y-1)/2] \label{TThc2_1} \eeq
\beq (T_{T,hc,L_y,L_y-1,2})_{L_y+2j-1,L_y+2j-1}=1 \quad {\rm for} \quad 1
\le j \le [L_y/2] \label{TThc2ya} \eeq
\beq (T_{T,hc,L_y,L_y-1,2})_{L_y+2j,L_y+2j}=y \quad {\rm for} \quad 1 \le
j \le [(L_y-1)/2] \label{TThc2yb} \eeq
\beq
(T_{T,hc,L_y,L_y-1,2})_{2j-1,L_y+2j-1}=(T_{T,hc,L_y,L_y-1,2})_{2j,L_y+2j-1}
=\frac{1}{y-1} \quad {\rm for} \quad 1 \le j \le [L_y/2]
\label{TThc2wa} \eeq
\beq (T_{T,hc,L_y,L_y-1,2})_{2j,L_y+2j}=(T_{T,hc,L_y,L_y-1,2})_{2j+1,L_y+2j}
=\frac{y}{y-1} \quad {\rm for} \quad 1 \le j \le [(L_y-1)/2]
\label{TThc2wb} \eeq
\beq (T_{T,hc,L_y,L_y-1,2})_{L_y+2j,2j}=(T_{T,hc,L_y,L_y-1,1})_{L_y+2j,2j+1}
=y-1 \quad {\rm for} \quad 1 \le j \le [(L_y-1)/2] \label{TThc2v}
\eeq
with all other elements equal to zero.

For general $x$ and $y$, $T_{T,hc,L_y,L_y-1}$ has rank equal to its dimension,
$2L_y-1$.  In the special case $y=0$ (chromatic polynomial), the rank is
reduced to $(3L_y-1)/2$ for odd $L_y$ and to $(3/2)L_y-1$ for even $L_y$ (and
may be reduced further for particular $x$).  In the special case $x=0$ (flow
polynomial), the rank is reduced to $L_y-1$ (and may be reduced further for
particular $y$).  These results are in accord with our previous general
findings that \cite{cf} $n_P(sq,L_y,d)=n_P(tri,L_y,d)$ and \cite{hca} 
$n_F(sq,L_y,d)=n_F(hc,L_y,d)$. 

We illustrate these general formulas for the cases $L_y=2,3,4$. 
For this purpose we use the abbreviations $x_1=1+x$, $x_2=2+x$
and $r_d=y/(y-1)$ as before.
\beq T_{T,hc,2,1,1} = \left( \begin{array}{ccc}
    x_1 & 1  &  r_d \\
    1  & x_1 &  r_d \\
    v  & v  &  y \end{array} \right )
    = T_{T,sq,2,1} \ , \qquad
    T_{T,hc,2,1,2} = \left( \begin{array}{ccc}
    x & 0 & 1/v \\
    0 & x & 1/v \\
    0 & 0 & 1 \end{array} \right )
\label{TThc21parts} \eeq
\beq T_{T,hc,3,2,1} = \left( \begin{array}{ccccc}
    x_1 & 1  & 0 & r_d  & 0 \\
    1  & x_1 & 0 & r_d  & 1/v \\
    0  & 0  & x & 0   & 1/v \\
    v  & v  & 0 & y  & 0 \\
    0  & 0  & 0 & 0  & 1 \end{array} \right )
    \ , \qquad
    T_{T,hc,3,2,2} = \left( \begin{array}{ccccc}
    x & 0  & 0  & 1/v & 0 \\
    0 & x_1 & 1  & 1/v & r_d \\
    0 & 1  & x_1 & 0   & r_d \\
    0 & 0  & 0  & 1   & 0 \\
    0 & v  & v  & 0   & y \end{array} \right )
\label{TThc32parts} \eeq
\beqs T_{T,hc,4,3,1} & = & \left( \begin{array}{ccccccc}
    x_1 & 1  & 0  & 0  & r_d & 0  & 0  \\
    1  & x_1 & 0  & 0  & r_d & 1/v& 0  \\
    0  & 0  & x_1 & 1  & 0  & 1/v& r_d \\
    0  & 0  & 1  & x_1 & 0  & 0  & r_d \\
    v  & v  & 0  & 0  & y  & 0  & 0  \\
    0  & 0  & 0  & 0  & 0  & 1  & 0  \\
    0  & 0  & v  & v  & 0  & 0  & y \end{array} \right )
    \cr\cr
& & \cr\cr
    T_{T,hc,4,3,2} & = & \left( \begin{array}{ccccccc}
    x & 0  & 0  & 0 & 1/v& 0  & 0  \\
    0 & x_1 & 1  & 0 & 1/v& r_d & 0  \\
    0 & 1  & x_1 & 0 & 0  & r_d & 1/v\\
    0 & 0  & 0  & x & 0  & 0  & 1/v\\
    0 & 0  & 0  & 0 & 1  & 0  & 0  \\
    0 & v  & v  & 0 & 0  & y  & 0  \\
    0 & 0  & 0  & 0 & 0  & 0  & 1 \end{array} \right )
\label{TThc43parts} \eeqs

The $T_{X,hc,L_y,L_y-1}$, $X=T,Z$ are obtained via eq. (\ref{transfermatrix})
from these auxiliary matrices.  For example, 
\beq
T_{T,hc,2,1}= \left( \begin{array}{ccc}
   1+x+x^2   & 1+x      &  (1+x)y(y-1)^{-1}    \\
    1+x      & 1+x+x^2  &   (1+x)y(y-1)^{-1}   \\  
    y-1      & y-1      &    y           \end{array} \right )
\label{TThc21}
\eeq
or equivalently, 
\beq
T_{Z,hc,2,1}= \left( \begin{array}{ccc}
v^2(3v^2+3vq+q^2)    &  v^3(2v+q)          &  v^2(2v+q)(1+v)   \\
v^3(2v+q)            & v^2(3v^2+3vq+q^2)   &  v^2(2v+q)(1+v)   \\
v^5                  & v^5                 &  v^4(1+v) \end{array} \right )
\label{TZhc21}
\eeq

We find that in general, for degree $d=L_y-1$, one of the $2L_y-1$ eigenvalues,
say that for $j=2L_y-1$, has a particularly simple expression:
\beq
\lambda_{T,hc,L_y,d=L_y-1,j=2L_y-1}=x^2
\label{lamtut_hcdm1}
\eeq
or equivalently, 
\beq
\lambda_{Z,hc,L_y,d=L_y-1,j=2L_y-1}=v^{2(L_y-1)}(v+q)^2 \ . 
\label{lam_hcdm1}
\eeq
The expressions for the other $2L_y$ eigenvalues are, in general, more
complicated.

\section{General Results for Cyclic Self-Dual Square-Lattice Strips}

In this section we consider the Potts model for families of self-dual strip
graphs of the square lattice with fixed width $L_y$ and arbitrarily great
length $L_x$, having periodic longitudinal boundary conditions, such that all
vertices on one side of the strip, which we take to be the upper side are
joined by edges to a single external vertex.  A strip graph of this type will
be denoted generically as $G_D$ and, in more detail, as $G_D(L_y\times L_x)$.
The family of $G_D$ graphs is planar and self-dual.  In general, the graph
$G_D(L_y \times L_x)$ has $n \equiv |V|= L_xL_y+1$ vertices, equal to the
number of faces, $f$.  One motivation for considering the $G_D$ strip graphs is
that they exhibit, for any $L_y$, the self-duality property of the infinite
square lattice so that, by (\ref{tdual}),
\beq
T(G_D,x,y)=T(G_D,y,x) \ .
\label{tsdgsym}
\eeq
Equivalently, by (\ref{zdual}), aside from a prefactor, the partition function
is invariant under $v \to q/v$.  This important duality property is used to
derive the formula for the transition temperature of the $q$-state Potts
ferromagnet, which occurs at the self-dual point $v=\sqrt{q}$.  

In Ref. \cite{dg} we gave the general form for $Z(G_D,L_y \times L_x,q,v)$
which, in our current notation with $m$ given in terms of $L_x$ by
eq. (\ref{lxm}), is 
\beq
Z(G_D,L_y \times L_x,q,v) = \sum_{d=1}^{L_y+1} \kappa^{(d)}
Tr[(T_{Z,G_D,L_y,d})^m]
\label{zsdgtran}
\eeq
or equivalently 
\beq
T(G_D,L_y \times L_x,x,y) = q^{-1} \sum_{d=1}^{L_y+1} \kappa^{(d)}
Tr[(T_{T,G_D,L_y,d})^m]
\label{tsdgtran}
\eeq
where the factor of $q^{-1}$ arising from a prefactor of $1/(x-1)$, and 
\beqs
\kappa^{(d)} & = & \sqrt{q} \ U_{2d-1} \Big ( \frac{\sqrt{q}}{2} \Big ) \cr\cr
& = & \sum_{j=0}^{d-1} (-1)^j { 2d-1-j \choose j} q^{d-j} \ . 
\label{kappad}
\eeqs
The first few of these coefficients are 
\beq
\kappa^{(1)}=q, \quad \kappa^{(2)}=q(q-2), \quad \kappa^{(3)}=
q(q-1)(q-3), \quad \kappa^{(4)}=q(q-2)(q^2-4q+2) \ . 
\label{kappabarn}
\eeq
One has 
\beq
\kappa^{(d)} = \prod_{k=1}^d (q - s_{d,k})
\label{kappafactors}
\eeq
where
\beq
s_{d,k} = 4\cos^2 \bigg ( \frac{\pi k}{2d} \bigg ) \quad {\rm for} \ 
 k=1,2,..d
\label{skd}
\eeq
and
\beq
\kappa^{(d)}=c^{(d)}+c^{(d-1)} \quad {\rm for} \ d = 1,2,...
\label{kappacd}
\eeq

The dimension $dim(T_{Z,G_D,L_y,d})=n_Z(G_D,L_y,d)$ was calculated in Ref.
\cite{dg}; two relevant results are
\beq
n_Z(G_D,L_y,L_y)=2L_y \ , \quad n_Z(G_D,L_y,L_y+1)=1 \ . 
\label{nzgdvalues}
\eeq

Combining eq. (\ref{tsdgsym}), (\ref{tsdgtran}), and the fact that
$\kappa^{(d)}$ is a function of $q$ which, in turn, is a symmetric function
under $x \leftrightarrow y$ (cf. eq. (\ref{qxy})), it follows that the traces
of the $m$'th powers of the transfer matrices themselves and their eigenvalues
also have this symmetry:
\beq
Tr[(T_{G_D,L_y,d}(x,y))^m] = Tr[(T_{G_D,L_y,d}(y,x))^m] 
\label{tsdgdual}
\eeq
\beq
\lambda_{T,G_D,L_y,d}(x,y) = \lambda_{T,G_D,L_y,d}(y,x)  \ . 
\label{lamgdsym}
\eeq
where we have explicitly indicated the functional dependence of the transfer
matrices and eigenvalues on the variables $x$ and $y$.

\subsection{Determinants}

We find
\beq
det(T_{T,G_D,L_y,d}) = (xy)^{L_y n_Z(G_D,L_y-1,d)} 
\label{detTTsdgld}
\eeq
for the full range $1 \le d \le L_y+1$. 

\subsection{Traces}

For the total trace, defined as 
\beq
Tr(T_{T,G_D,L_y}) = q^{-1}\sum_{d=1}^{L_y+1} \kappa^{(d)} Tr(T_{T,G_D,L_y,d})
\label{deftracegd}
\eeq
we find 
\beq
Tr(T_{T,G_D,L_y}) = (xy)^{L_y} \ . 
\label{tracegd}
\eeq

\subsection{Eigenvalue for $d=L_y+1$ for $\Lambda=G_D$}

For $d=L_y+1$, the transfer matrix $T_{T,G_D,L_y,d}$ reduces to a scalar,
namely 
\beq
T_{T,G_D,L_y,L_y+1}=\lambda_{T,G_D,L_y,L_y+1}=1 \ . 
\label{lamtutsdgly}
\eeq

\subsection{Transfer Matrix for $d=L_y$, $\Lambda=G_D$}

The transfer matrix $T_{T,G_D,L_y,L_y}$ has dimension $2L_y$.  We obtain the
following general formula, which we write for the Tutte polynomial, since it
has a somewhat simpler form.
\beq
(T_{T,G_D,L_y,L_y})_{1,1}=1+x 
\label{TTsdgx1}
\eeq
\beq
(T_{T,G_D,L_y,L_y})_{j,j}=2+x \quad {\rm for} \quad
L_y \ge 2 \ \ {\rm and} \ \ 2 \le j \le L_y
\label{TTsdgx2}
\eeq
\beq
(T_{T,G_D,L_y,L_y})_{j,j+1} = (T_{T,G_D,L_y,L_y})_{j+1,j} = 1 
\quad {\rm for} \quad 1 \le j \le L_y-1
\label{TTsdg1a}
\eeq
\beq
(T_{T,G_D,L_y,L_y})_{j,j}=y \quad {\rm for} \quad L_y+1 \le j \le 2L_y
\label{TTsdgy}
\eeq
\beq
(T_{T,G_D,L_y,L_y})_{j,L_y+j}=\frac{y}{y-1} \quad {\rm for} \quad 
1 \le j \le L_y
\label{TTsdgwa}
\eeq
\beq
(T_{T,G_D,L_y,L_y})_{j+1,L_y+j}=\frac{y}{y-1} \quad {\rm for} \quad 
1 \le j \le L_y-1 
\label{TTsdgwb}
\eeq
\beq
(T_{T,G_D,L_y,L_y})_{L_y+j,j}= y-1 \quad {\rm for} \quad 1 \le j \le L_y
\label{TTsdgva}
\eeq
\beq
(T_{T,G_D,L_y,L_y})_{L_y+j,j+1}= y-1 \quad {\rm for} \quad 1 \le j \le L_y-1
\label{TTsdgvb}
\eeq
with all other elements equal to zero.  The elements of $T_{Z,G_D,L_y,L_y}$
are given by the relation (\ref{tztt}) so that 
$\lambda_{Z,G_D,L_y,d}=v^{L_y}\lambda_{T,G_D,L_y,d}$. 

For general $x$ and $y$, $T_{T,G_D,L_y,L_y}$ has rank equal to its dimension,
$2L_y$.  In the special cases $y=0$ and $x=0$, which yield the chromatic and
flow polynomials, respectively, the rank is reduced to $L_y$ and remains equal
to $L_y$ if both $x$ and $y$ are zero.  

We illustrate these general formulas for the cases $L_y=1,2,3$:
\beq
T_{T,G_D,1,1} = \left( \begin{array}{cc}
    x_1 &    r_d \\
    v   &    y   \end{array} \right )
\label{TTsdg11}
\eeq
\beq
T_{T,G_D,2,2} = \left( \begin{array}{cccc}
    x_1  & 1   &  r_d  &  0  \\
    1    & x_2 &  r_d  & r_d \\
    v    & v   &   y   & 0   \\
    0    & v   & 0     & y   \end{array} \right )
\label{TTsdg22}
\eeq
\beq
T_{T,G_D,3,3} = \left( \begin{array}{cccccc}
    x_1  & 1   &  0  & r_d  & 0    & 0   \\
    1    & x_2 &  1  & r_d  & r_d  & 0   \\
    0    & 1   & x_2 & 0    & r_d  & r_d \\
    v    & v   & 0   & y    & 0    & 0   \\
    0    & v   & v   & 0    & y    & 0   \\
    0    & 0   & v   & 0    & 0    & y  \end{array} \right )
\label{TTsdg33}
\eeq
Thus, in general, the upper left-hand submatrix has a main diagonal with the
first entry equal to $x_1$ and the other entries equal to $x_2$.  Adjacent to
this main diagonal are two diagonals whose entries are 1, and the rest of the
submatrix is comprised of triangular regions filled with 0's.  In the upper
right-hand submatrix the main diagonal and the adjacent diagonal below it have
entries equal to $r_d$, and the other entries are zero.  In the lower left-hand
submatrix the main diagonal and the adjacent diagonal above it have entries
equal to $v$, and the other entries are zero.  And finally, in the right-hand
lower submatrix the entries on the main diagonal are equal to $y$ and the rest
of this submatrix is made up of triangular regions of 0's.  For the lowest
values $L_y=1,2$, some of these parts, such as the triangular regions of zeros,
are absent. 

Using eq. (\ref{tztt}), it is straightforward to obtain the
$T_{Z,G_D,L_y,L_y}$ from these $T_{T,G_D,L_y,L_y}$ matrices.  For example, 
\beq
T_{Z,G_D,1,1}= \left( \begin{array}{cc}
 q+2v  & 1+v         \\
  v^2  & v(1+v)      \end{array} \right )
\label{TZsdg11}
\eeq

As corollaries of our general result for $T_{T,G_D,L_y,L_y}$ we calculate the
determinant and trace:
\beq
det(T_{T,G_D,L_y,L_y}) = (xy)^{L_y}
\label{detTsdglyly}
\eeq
which is the $d=L_y$ special case of (\ref{detTTsdgld}), and 
\beq
Tr(T_{T,G_D,L_y,L_y}) = L_y(x+y) + 2L_y-1 \ . 
\label{traceTsdglyly}
\eeq

\section{Some Illustrative Calculations}

\subsection{Square-Lattice Strip, $L_{\lowercase{y}}=2$}

The Potts model partition function $Z(sq,L_y \times m,BC,q,v)$ and Tutte
polynomial $T(sq,L_y \times m,BC,q,v)$ (BC = boundary conditions) were
calculated for the cyclic and M\"obius strips of the square lattice of width
$L_y=2$ in Ref. \cite{a}.  We express the results here in terms of transfer
matrices $T_{X,sq,2,d}$, $X=Z,T$, via eqs. (\ref{zgsum_transfer}) and
(\ref{tgsum_transfer}):
\beq
T_{T,sq,2,0} = \left( \begin{array}{cc}
    1+x+x^2 & x_1 r_d \\
    v       & y       \end{array} \right )
\label{TTsq20}
\eeq
\beq
T_{Z,sq,2,0} = \left( \begin{array}{cc}
    q^2+3qv+3v^2 & (q+2v)(1+v) \\
    v^3          & v^2(1+v) \end{array} \right )
\label{TZsq20}
\eeq
The matrices $T_{T,sq,2,1}$ and $T_{T,sq,2,2}$ have been given above.  The
corresponding results for the $L_y=2$ M\"obius strip follow from our general
formulas also given above.

\subsection{Triangular-Lattice Strip, $L_{\lowercase{y}}=2$}

We illustrate our results for the $L_y=2$ cyclic strip of the triangular
lattice. We obtain 
\beq
T_{T,tri,2,0} = \left( \begin{array}{cc}
    (1+x)^2+y    & (1+x+y)r_d \\
    (y-1)(1+x+y) & y(1+y)      \end{array} \right )
\label{TTtri20}
\eeq
or equivalently, 
\beq
T_{Z,tri,2,0} = \left( \begin{array}{cc}
    5v^2+4qv+q^2+v^3  & (q+3v+v^2)(1+v) \\
    v^2(q+3v+v^2)     & v^2(2+v)(1+v)      \end{array} \right )
\label{TZtri20}
\eeq
The matrices $T_{T,tri,2,1}$ and $T_{T,tri,2,2}$ were given above. 

\subsection{Honeycomb-lattice Strip, $L_{\lowercase{y}}=2$}

For the $L_y=2$ cyclic strip of the honeycomb lattice we calculate
\beq
T_{T,hc,2,0}= \left( \begin{array}{cc}
     \sum_{j=0}^4 x^j   &   \ \ (1+x)(1+x^2)y(y-1)^{-1}   \\
        y-1       &   \ \ y  \end{array} \right )
\label{TThc20}
\eeq
or equivalently, 
\beq
T_{Z,hc,2,0}= \left( \begin{array}{cc}
          h_{11}   &  h_{12}      \\
          v^5       &   v^4(1+v)  \end{array} \right )
\label{TZhc20}
\eeq
\beq
h_{11}=5v^4+10v^3q+10v^2q^2+5vq^3+q^4
\label{th11}
\eeq
\beq
h_{12}=(2v+q)(2v^2+2vq+q^2)(1+v)
\label{th12}
\eeq
The other matrices relevant for the strip were given above. 

\bigskip

Acknowledgments: We thank J. Salas for discussions on transfer matrix methods
during the work for Refs. \cite{ts,tt}, A. Sokal for related discussions, and
N. Biggs for discussion on sieve methods.  The research of R.S. was partially
supported by the NSF grant PHY-00-98527.  The research of S.C.C. was partially
supported by the Nishina and Inoue Foundations, and he thanks Prof. M. Suzuki
for further support.  The NCTS Taipei address for S.C.C. applies after April
12, 2004.

\section{Appendix}

In this appendix we include tables of the numbers $\Delta n_Z(sq,L_y,d)$ and
$n_Z(sq,L_y,d,\pm)$ discussed in the text.

\begin{table}
\caption{\footnotesize{Table of $\Delta n_Z(sq,L_y,d)$ for strips of the square
lattice. Blank entries are zero. The last entry for each value of $L_y$ is the
total number of partitions with self-reflection symmetry.}}
\begin{center}
\begin{tabular}{|c|c|c|c|c|c|c|c|c|c|c|c|c|}
$L_y$ & 0 & 1 & 2 & 3 & 4 & 5 & 6 & 7 & 8 & 9 & 10 & $\Delta
N_{Z,L_y}$ \\ \hline\hline 
1  & 1  & 1  &    &    &    &    &   &   &   &   & & 2  \\ \hline 
2  & 2  & 1  & 1  &    &    &    &   &   &   &   & & 4  \\ \hline 
3  & 3  & 3  & 1  & 1  &    &    &   &   &   &   & & 8  \\ \hline 
4  & 6  & 4  & 4  & 1  & 1  &    &   &   &   &   & & 16 \\ \hline 
5  &10  &10  & 5  & 5  & 1  & 1  &   &   &   &   & & 32 \\ \hline 
6  &20  &15  &15  & 6  & 6  & 1  & 1 &   &   &   & & 64 \\ \hline 
7  &35  &35  &21  &21  & 7  & 7  & 1 & 1 &   &   & & 128 \\ \hline 
8  &70  &56  &56  &28  &28  & 8  & 8 & 1 & 1 &   & & 256 \\ \hline 
9  &126 &126 &84  &84  &36  &36  & 9 & 9 & 1 & 1 & & 512 \\ \hline
10 &252 &210 &210 &120 &120 &45 &45 &10 &10 & 1 & 1&1024 \\
\end{tabular}
\end{center}
\label{nzpmtable}
\end{table}

\begin{table}
\caption{\footnotesize{Table of numbers $n_Z(sq,L_y,d,\pm)$ for strips
of the square lattice.  For each $L_y$ value, the entries in the
first and second lines are $n_Z(sq,L_y,d,+)$ and $n_Z(sq,L_y,d,-)$,
respectively. Blank entries are zero. The last entry for each
value of $L_y$ is the total $N_{Z,L_y,\lambda}$.}}
\begin{center}
\begin{tabular}{|c|c|c|c|c|c|c|c|c|c|c|c|c|}
$L_y \ (d,+) $
   & $0,+$ & $1,+$ & $2,+$ & $3,+$ & $4,+$ & $5,+$ & $6,+$ & $7,+$ & $8,+$ & $9,+$ & $10,+$ & \\
\quad \ $(d,-) $
   & $0,-$ & $1,-$ & $2,-$ & $3,-$ & $4,-$ & $5,-$ & $6,-$ & $7,-$ & $8,-$ & $9,-$ & $10,-$
& $N_{Z,L_y,\lambda}$ \\ \hline\hline
2  & 2    & 2     & 1     &       &       &      &      &     &    &    &   &       \\
   &      & 1     &       &       &       &      &      &     &    &    &   & 6     \\ \hline
3  & 4    & 6     & 3     & 1     &       &      &      &     &    &    &   &       \\
   & 1    & 3     & 2     &       &       &      &      &     &    &    &   & 20    \\ \hline
4  & 10   & 16    & 12    & 4     & 1     &      &      &     &    &    &   &       \\
   & 4    & 12    & 8     & 3     &       &      &      &     &    &    &   & 70    \\ \hline
5  & 26   & 50    & 40    & 20    & 5     & 1    &      &     &    &    &   &       \\
   & 16   & 40    & 35    & 15    & 4     &      &      &     &    &    &   & 252   \\ \hline
6  & 76   & 156   & 145   & 80    & 30    & 6    & 1    &     &    &    &   &       \\
   & 56   & 141   & 130   & 74    & 24    & 5    &      &     &    &    &   & 924   \\ \hline
7  & 232  & 518   & 511   & 329   & 140   & 42   & 7    & 1   &    &    &   &       \\
   & 197  & 483   & 490   & 308   & 133   & 35   & 6    &     &    &    &   & 3432  \\ \hline
8  & 750  & 1744  & 1848  & 1288  & 644   & 224  & 56   & 8   & 1  &    &   &       \\
   & 680  & 1688  & 1792  & 1260  & 616   & 216  & 48   & 7   &    &    &   & 12870 \\ \hline
9  & 2494 & 6030  & 6672  & 5040  & 2772  & 1140 & 336  & 72  & 9  & 1  &   &       \\
   & 2368 & 5904  & 6588  & 4956  & 2736  & 1104 & 327  & 63  & 8  &    &   & 48620 \\ \hline
10 & 8524 & 21100 & 24330 & 19440 & 11688 & 5352 & 1875 & 480 & 90 & 10 & 1 &       \\
   & 8272 & 20890 & 24120 & 19320 & 11568 & 5307 & 1830 & 470 & 80 & 9  &   &184756 \\ 
\end{tabular}
\end{center}
\label{ntctable}
\end{table}

\vfill
\eject

\newpage

\begin{thebibliography}{99}

\bibitem{potts}
R. B. Potts, Proc. Camb. Phil. Soc. {\bf 48} (1952) 106.

\bibitem{wurev}
F. Y. Wu, Rev. Mod. Phys. {\bf 54} (1982) 235.

\bibitem{kf}
P. W. Kasteleyn and C. M. Fortuin, J. Phys. Soc. Jpn. (Suppl.) {\bf 26} (1969)
11. 

\bibitem{fk}
C. M. Fortuin and P. W. Kasteleyn, Physica {\bf 57} (1972) 536.

\bibitem{tutte1}
W. T. Tutte, Can. J. Math. {\bf 6} (1954) 80.

\bibitem{tutte2}
W. T. Tutte, J. Combin. Theory {\bf 2} (1967) 301.

\bibitem{tutte3}
W. T. Tutte, ``Chromials'', in Lecture Notes in Math. v. 411
(1974) 243; {\it Graph Theory}, vol. 21 of Encyclopedia of
Mathematics and Applications (Addison-Wesley, Menlo Park, 1984).

\bibitem{bbook}
N. L. Biggs, {\it Algebraic Graph Theory} (Cambridge
Univ. Press, Cambridge, 1st ed. 1974, 2nd ed. 1993).

\bibitem{rtrev}
R. C. Read and W. T. Tutte, ``Chromatic Polynomials'',
in {\it Selected Topics in Graph Theory, 3}, (Academic Press, New York, 1988),
p. 15.

\bibitem{boll}
B. Bollob\'as, {\it Modern Graph Theory} (Springer, New York, 1998).

\bibitem{a}
R. Shrock, Physica A {\bf 283} (2000) 388.

\bibitem{cf}
S.-C. Chang and R. Shrock, Physica A {\bf 296} (2001) 131. 

\bibitem{s3a}
S.-C. Chang and R. Shrock, Physica A {\bf 296} (2001) 234. 

\bibitem{ta}
S.-C. Chang and R. Shrock, Physica A {\bf 286} (2001) 189. 

\bibitem{hca}
S.-C. Chang and R. Shrock, Physica A {\bf 296} (2001) 183. 

\bibitem{jz}
S.-C. Chang and R. Shrock,  Physica A {\bf 301} (2001) 196.
            
\bibitem{dg}
S.-C. Chang and R. Shrock, Physica A {\bf 301} (2001) 301. 

\bibitem{sdg}
S.-C. Chang and R. Shrock, Phys. Rev. E {\bf 64} (2001) 066116. 

\bibitem{ka}
S.-C. Chang and R. Shrock, Int. J. Mod. Phys. B {\bf 15} (2001) 443. 

\bibitem{ka3}
S.-C. Chang and R. Shrock, in {\it Proc. of the CRM Workshop on Tutte
Polynomials, 2001}, Advances in Applied. Math. {\bf 32} (2004) 44.

\bibitem{s5}
S.-C. Chang and R. Shrock, Physica A {\bf 316} (2002) 335.

\bibitem{bds}
N. L. Biggs, R. M. Damerell, and D. A. Sands, J. Combin. Theory
B {\bf 12} (1972) 123.

\bibitem{bm}
N. L. Biggs and G. H. Meredith, J. Combin. Theory B {\bf 20} (1976) 5.

\bibitem{baxter86}
R. J. Baxter, J. Phys. A {\bf 19} (1986) 2821.

\bibitem{baxter87}
R. J. Baxter, J. Phys. A {\bf 20} (1987) 5241.

\bibitem{matmeth}
N. L. Biggs, J. Combin. Theory B {\bf 82} (2001) 19.

\bibitem{matmeth2}
N. L. Biggs, Bull. London Math. Soc. {\bf 34} (2002) 129.

\bibitem{matmeth3}
N. L. Biggs, London School of Economics Centre for Discrete and Applicable
Mathematics, report LSE-CDAM-2000-04.

\bibitem{sqtran}
J. Salas and A. Sokal, J. Stat. Phys. {\bf 104} (2001) 609.

\bibitem{ts}
S.-C. Chang, J. Salas, and R. Shrock, J. Stat. Phys. {\bf 107} (2002) 1207.

\bibitem{tt}
S.-C. Chang, J. Jacobsen, J. Salas, and R. Shrock,
J. Stat. Phys. {\bf 114} (2004) 763.

\bibitem{ks}
H. Kluepfel and R. Shrock, YITP-99-32; H. Kluepfel, Stony Brook
thesis (July, 1999).

\bibitem{ck}
S.-Y. Kim and R. Creswick, Physical Review E {\bf 63} (2001) 066107. 

\bibitem{pt}
S.-C. Chang and R. Shrock, cond-mat/0404373. 

\bibitem{triform}
See, e.g., P. Lancaster and M. Tismenetsky, {\it The Theory of Matrices}
(Academic Press, New York, 1985).

\bibitem{pm}
R. Shrock, Phys. Lett. A {\bf 261} (1999) 57.

\bibitem{uspensky}{J. V. Uspensky, {\it Theory of Equations}
(McGraw-Hill, NY 1948), 264.}

\bibitem{f}
S.-C. Chang and R. Shrock, J. Stat. Phys., {\bf 112} (2003) 815.

\bibitem{tor}
S.-C. Chang and R. Shrock, Physica A {\bf 292} (2001) 307.

\end{thebibliography}
\end{document}